\documentclass[prx,superscriptaddress,twocolumn,amsmath,amssymb,nofootinbib]{revtex4-2}

\usepackage{mathtools}
\DeclarePairedDelimiter\abs{\lvert}{\rvert}

\usepackage{hyperref}
\usepackage{graphicx}
\usepackage{natbib}

\usepackage{diffcoeff}
\usepackage{braket}
\usepackage[low-sup]{subdepth}
\usepackage{mleftright}
\mleftright
\usepackage{enumitem}

\usepackage[dvipsnames]{xcolor}

\newcommand\Loc{\mathrm{loc}}
\newcommand\Int{\mathrm{int}}
\newcommand\A{\mathrm A}
\newcommand\B{\mathrm B}
\renewcommand\phi\varphi
\newcommand\e{\mathrm e}
\newcommand\ii{\mathrm i}
\newcommand\M{\mathrm M}
\renewcommand{\rho}{\varrho}
\DeclareMathOperator{\sgn}{sgn}
\newcommand\be{\begin{equation}}
\newcommand\ee{\end{equation}}

\newcommand\ce\varepsilon
\DeclareMathOperator\Tr{Tr}

\begin{document}

\title{Many-body quantum vacuum fluctuation engines}
\date{\today}

\author{\'Etienne Jussiau}
\email{etienne.jussiau@gmail.com}
\affiliation{Department of Physics and Astronomy, University of Rochester, Rochester, NY 14627, USA}
\affiliation{Institute for Quantum Studies, Chapman University, Orange, CA 92866, USA}
\author{L\'ea Bresque}
\affiliation{Universit\'e Grenoble Alpes, CNRS, Grenoble INP, Institut N\'eel, 38000 Grenoble, France}
\author{Alexia Auff\`eves}
\affiliation{MajuLab, CNRS–UCA-SU-NUS-NTU International Joint Research Laboratory}
\affiliation{Centre for Quantum Technologies, National University of Singapore, 117543 Singapore, Singapore}
\author{Kater W. Murch}
\affiliation{Department of Physics, Washington University, St. Louis, Missouri 63130, USA}
\author{Andrew N. Jordan}
\email{jordan@chapman.edu}
\affiliation{Institute for Quantum Studies, Chapman University, Orange, CA 92866, USA}
\affiliation{Department of Physics and Astronomy, University of Rochester, Rochester, NY 14627, USA}

\begin{abstract}
We propose a many-body quantum engine powered by the energy difference between the entangled ground state of the interacting system and local separable states. Performing local energy measurements on an interacting many-body system can produce excited states from which work can be extracted via local feedback operations. These measurements reveal the quantum vacuum fluctuations of the global ground state in the local basis and provide the energy required to run the engine. The reset part of the engine cycle is particularly simple: The interacting many-body system is coupled to a cold bath and allowed to relax to its entangled ground state. We illustrate our proposal on two types of many-body systems: a chain of coupled qubits and coupled harmonic oscillator networks. These models faithfully represent fermionic and bosonic excitations, respectively. In both cases, analytical results for the work output (average value and standard deviation) and efficiency of the engine are derived. We prove the efficiency is controlled by the ``local entanglement gap''---the energy difference between the many-body ground state and the lowest energy eigenstate of the local Hamiltonian. In all the examples analyzed in this work, for a large number of coupled subsystems, the average work output scales linearly or faster and dominates over fluctuations, while the efficiency limits to a constant. In the qubit chain case, we highlight the impact of a quantum phase transition on the engine's performance as work and efficiency sharply increase at the critical point. In the case of a one-dimensional oscillator chain, we show the efficiency approaches unity as the number of coupled oscillators increases, even at finite work output.
\end{abstract}

\maketitle

\section{Introduction}

A long-standing dream has been to harness quantum vacuum fluctuations in the design of new quantum machines. Pioneering studies have examined the possibility of using the Casimir effect to extract energy from the quantum vacuum and store it in a battery~\cite{PhysRevB.30.1700,PhysRevE.48.1562}. More recent works have analyzed the role of quantum vacuum fluctuations in various quantum technology platforms: It has been shown that zero-point fluctuations stemming from bosonic environments permit the rectification of electrical current, producing work~\cite{PhysRevB.92.125306}, and a superconducting circuit-based thermal engine relying on the presence of zero-point fluctuations of a microwave cavity has been proposed~\cite{PhysRevB.93.041418}. These works demonstrated that incorporating vacuum fluctuations in the design principles of engines and batteries is a scientifically sound and promising direction to pursue as quantum technology continues to develop.

Here we contribute to this endeavor by proposing an engine cycle whereby work is extracted from the quantum vacuum via local measurements in connection to recent works on measurement-powered machines. Indeed, advances in the emerging field of quantum energetics have led to the invention of different kinds of quantum machines powered not by the heat supplied by thermal baths, but by the ``quantum heat'' supplied by a measurement apparatus~\cite{elouard2017-1,elouard2017}. These so-called quantum measurement engines rectify the energy fluctuations associated with measurements made on observables that do not commute with the system Hamiltonian, and extract the energy in the form of useful work. Examples include qubit engines~\cite{elouard2017-1,elouard2017,yi2017,das2019,seah2020,cottet2017,naghiloo2018,masuyama2018}, quantum elevators~\cite{elouard2018,elouard2020,jordan2020}, the single-electron battery~\cite{elouard2018}, measurement-powered refrigeration~\cite{buffoni2019,dUrso2003,koski2015}, among other inventions~\cite{manikandan2022,ding2018,mohammady2017,koski2014,chida2017}. In particular, Refs.~\cite{elouard2017,elouard2018,jordan2020,manikandan2022} have demonstrated the possibility for measurement-powered engines to achieve unit efficiency without divergent temperatures.

A recent publication by some of us~\cite{bresque2021} puts forward a measurement engine where entanglement between two qubits is first generated via weak resonant interactions and then destroyed by local measurements. A local feedback operation is then used to extract useful work due to the qubits' energy detuning and reset the two-qubit system. This study introduces a qualitatively different type of engine that combines the concept of a quantum vacuum-powered engine with quantum measurements. Distinct from the previous work, our proposal relies on entanglement in the ground state which may occur in strongly interacting many-body systems. Thus, entanglement is a natural byproduct of relaxation to the ground state, and requires no tuned gate operations. It has been pointed out that local measurements of energy can serve as an entanglement witness for interacting systems in their ground state~\cite{jordan2004}, i.e. the probability of finding a particle to be in some excited state indicates the degree of entanglement with the rest of the environment. Here, we harness this insight to design an engine cycle using this quantum resource which is essentially free when the ground state is entangled.

The entangled ground state is, by definition, of smaller energy than any separable state. For a large class of interaction Hamiltonians, the entangled ground-state energy is also lower than the energy of any local eigenstate. This is notably the case when the expectation value of the interaction Hamiltonian in any local eigenstate is zero. These observations motivate our engine cycle where entanglement is first destroyed by local measurements, work is then extracted through local operations while the system is in its local eigenbasis, and entanglement is eventually recreated by letting the system relax to its interacting ground state. The key feature that enables the cycle to function is the energy gap between the ground state and any of the separable energy eigenstates of the local Hamiltonian. This is related to the concept of the ``entanglement gap,'' the energy difference between the ground state and the nearest energy of any separable state~\cite{PhysRevA.70.062113}. Here we focus on the smallest gap to the eigenstates of the local Hamiltonian, we coin as the \emph{local entanglement gap} hereafter. This local entanglement gap would naturally close if the local system Hamiltonian commutes with the total Hamiltonian, where local measurements are quantum non-demolition. Importantly, energy conservation imposes that the local entanglement gap is also the minimum amount of energy transferred during a local energy measurement on a system in its entangled ground state.

The paper is organized as follows: In Sec.~\ref{General}, we consider the engine protocol described above for a generic many-body system. This allows us to introduce the important concepts and derive the main results that will be used throughout this work. We then apply this formalism to many-qubit systems in Sec.~\ref{QBs}. We first consider the case of two coupled qubits in Sec.~\ref{2qb}, and then extend our analysis to an arbitrarily long chain in Sec.~\ref{Nqb}. The qubit chain can be mapped to a free-fermion model which enables us to obtain analytical results for the work output and efficiency of the many-qubit engine. Remarkably, the system undergoes a quantum phase transition, and we note that work and efficiency sharply increase at the critical point. In Sec.~\ref{bosonsec}, we address the bosonic case by considering systems of coupled oscillators that may be viewed as describing bosons such as photons or phonons. Similarly to the qubit case, we start by studying two coupled oscillators in Sec.~\ref{2osc} before considering an oscillator network in Sec.~\ref{Nosc}. Analytical expressions for the work and efficiency are derived with no restriction to one-dimensional systems. We give explicit results in one, two, and three dimensions for nearest-neighbor coupling, observing strikingly high efficiencies. We conclude in Sec.~\ref{ccl}.

\section{General formalism and results}
\label{General}

In this work, we are interested in extracting work out of the quantum vacuum through local measurements on entangled many-body states. We will then consider many-body systems whose Hamiltonian can be decomposed as follows:
\be
H=H_\Loc+H_\Int=\sum_jh_j+\sum_{j\ne k}g_{jk}\xi_j\xi_k.
\ee
The local Hamiltonian~$H_\Loc$ is the sum of one-body terms~$h_j$, where $h_j$ is the interaction-free Hamiltonian for subsystem~$j$. The interaction Hamiltonian~$H_\Int$ is the sum of two-body terms~$g_{jk}\xi_j\xi_k$ which describe the interaction between subsystems~$j$ and~$k$; $g_{jk}$ denotes the coupling amplitude, while $\xi_j$ and $\xi_k$ are operators acting on subsystems~$j$ and~$k$, respectively.

\begin{figure}
\centering
\includegraphics[width=\linewidth]{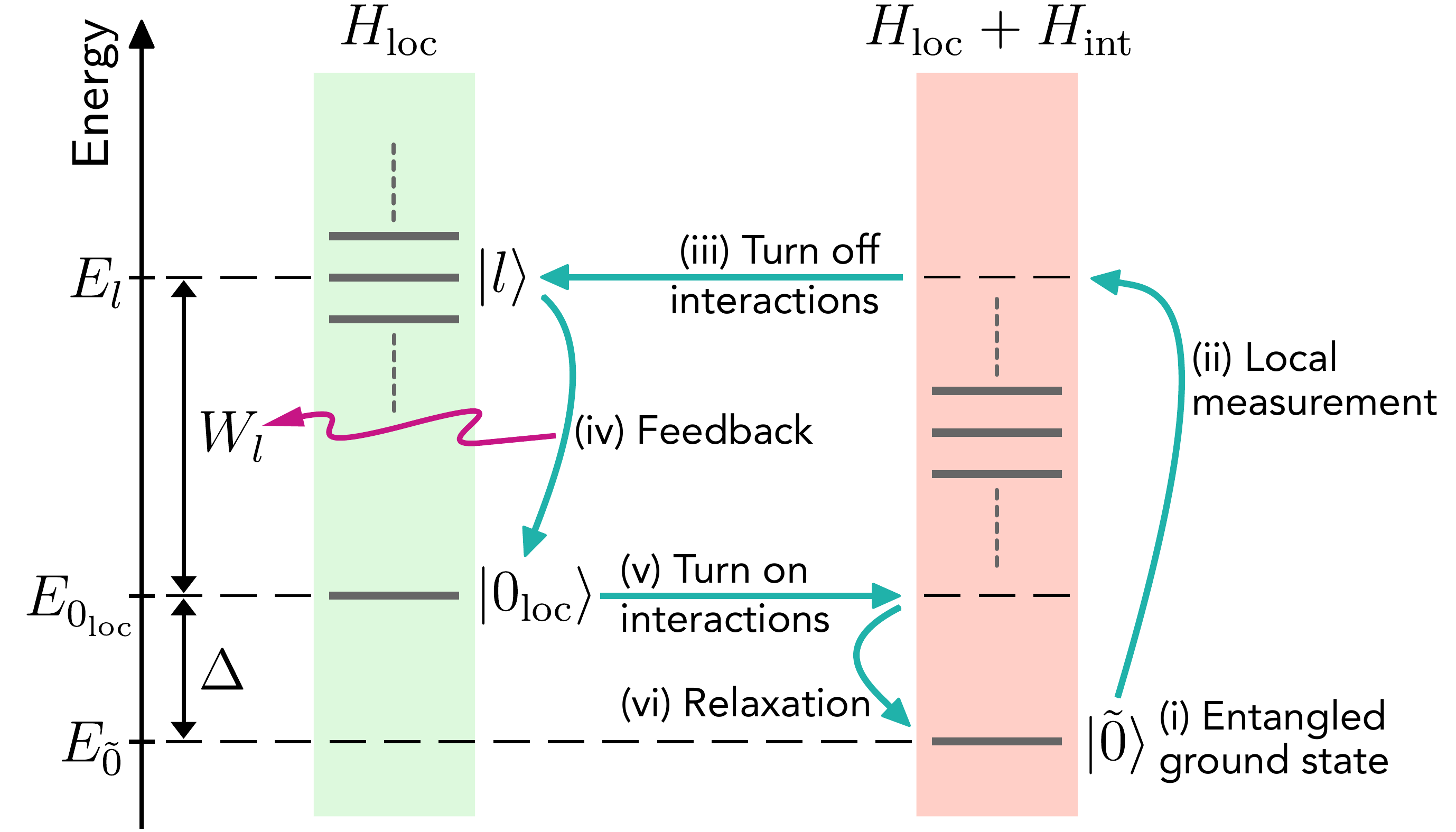}
\caption{Schematic depiction of the energy levels for the local (left) and interacting (right) Hamiltonians. The cyan arrows represent the energy changes throughout one possible realization of a cycle for the many-body vacuum fluctuation engine.}
\label{Lvls-Cycle}
\end{figure}

In what follows, we will write the states of the local eigenbasis as $\ket l=\otimes_j\ket{l_j}$, where the eigenstates of $h_j$ have been denoted by $\ket{l_j}$. In particular, $\ket{0_\Loc}=\otimes_j\ket{0_j}$ is the local ground state. We consider situations for which the ground state for the total Hamiltonian~$H$, denoted by $\ket{\tilde0}$ hereafter, is entangled, and it is therefore different from the local ground state: $\ket{\tilde0}\ne\ket{0_\Loc}$. We further require that the expectation value for any interaction operator is zero in the local eigenbasis: $\braket{l|\xi_j|l}=0$. This implies that the interaction Hamiltonian can be switched on and off with no energetic cost in the local eigenbasis since $\braket{l|H_\Int|l}=0$ for any local eigenstate~$\ket l$. As a consequence, the entangled ground state's energy~$E_{\tilde0}$ is necessarily lower than the local ground state's energy~$E_{0_\Loc}$:
\be
E_{\tilde0}=\braket{\tilde0|H|\tilde0}\le\braket{0_\Loc|H|0_\Loc}=\braket{0_\Loc|H_\Loc|0_\Loc}=E_{0_\Loc}.
\ee
Hereafter, the difference between these two energies will be denoted by $\Delta$ and referred to as the \emph{local entanglement gap},
\be
\Delta=E_{0_\Loc}-E_{\tilde0}\ge0.
\label{LocalEntanglementGap}
\ee
The notations introduced above for the energy levels of the local and interacting Hamiltonians can be visualized in Fig.~\ref{Lvls-Cycle}. 

To extract work out of the quantum vacuum, we break and restore the entangled ground state of our many-body system through the following cycle:
\begin{enumerate}[label=(\roman*),align=left]
\item Prepare the interacting many-body system in its entangled ground state.
\item Perform local measurements on each subsystem to project the many-body system onto the local eigenbasis.
\item Turn off interactions at no energetic cost.
\item Apply local control operations to bring any subsystem found in a local excited state to its local ground state in order to extract energy as useful work.
\item Turn interactions back on at no energetic cost.
\item Let the interacting many-body system relax to its entangled ground state.
\end{enumerate}
The variations of the many-body system's energy throughout the cycle are represented in Fig.~\ref{Lvls-Cycle}.

The probability to obtain the local eigenstate~$\ket l$ when performing local measurements on the interacting many-body system is given by $P_l=\abs{\braket{l|\tilde0}}^2$. The amount of work that can be extracted as the system is brought to its local ground state is then given by $W_l=E_l-E_{0_\Loc}$, where $E_l$ is the local eigenenergy associated with $\ket l$, i.e. $H_\Loc\ket l=E_l\ket l$. We deduce that the average work output for this cycle is given by the difference between the expectation value for the local Hamiltonian in the entangled ground state and the local ground-state energy:
\be
W=\sum_lP_lW_l=\braket{H_\Loc}_{\tilde0}-E_{0_\Loc}.
\label{Waverage}
\ee

The resource that enables this work extraction is the quantum heat provided by the measurement apparatus during the local measurements. Owing to energy conservation, when state~$\ket l$ is obtained as the result of local measurements on the interacting system in its ground state, the amount of energy transferred from the measurement apparatus to the system is $Q_l=E_l-E_{\tilde0}$. It is convenient to rewrite the heat in terms of the work output~$W_l$ and the local entanglement gap, $\Delta$, defined in Eq.~\eqref{LocalEntanglementGap}: $Q_l=E_l-E_{0_\Loc}+E_{0_\Loc}-E_{\tilde0}=W_l+\Delta$. This last equality can be interpreted as follows: Whatever its outcome, the local measurement destroys the entangled ground state which requires the measurement apparatus to supply an amount of energy at least equal to the local entanglement gap~$\Delta$. Moreover, if the result of the local measurement is a local excited state~$\ket l$, the additional amount of energy that must be provided by the measurement apparatus is equal to the amount of work~$W_l$ that will be extracted through local operations. The average heat supplied by the measurement apparatus can be straightforwardly obtained:
\be
Q=\sum_lP_lQ_l=W+\Delta=\braket{H_\Loc}_{\tilde0}-E_{\tilde0}=-\braket{H_\Int}_{\tilde0}.
\ee
In principle, the quantum heat is not the only energetic resource that must be supplied to the engine. Indeed, the energy cost of erasing the measurement outcome at the end of the cycle should also be taken into account. According to Landauer's principle~\cite{landauer1961,sagawa2009}, the minimal energy that must be supplied to a memory to erase a bit of information is $k_\mathrm BT\ln 2$, where $T$ is the temperature of the heat bath to which the memory is coupled. This energetic cost can then be made as small as desired by coupling the memory to a cold enough bath. In contrast, the quantum heat does not depend on external factors and can therefore be considered as the irreducible amount of energy that must be supplied to the engine. In the situation studied here, it is assumed that we have access to a very cold bath to relax the many-body system to its ground state at the end of the cycle. As a consequence, we will consider hereafter that the energetic cost of erasing the measurement outcome is negligible as compared with the quantum heat. The engine efficiency is then given by the ratio of the work output to the quantum heat:
\be
\eta=\frac WQ=\frac W{W+\Delta}=\frac{\braket{H_\Loc}_{\tilde0}-E_{0_\Loc}}{-\braket{H_\Int}_{\tilde0}}.
\ee

To conclude this section, we address the question of fluctuations between different realizations of the cycle. The work output's standard deviation~$\sigma$ is such that
\be
\sigma^2=\sum_lP_l(W_l-W)^2.
\ee
Similarly to the average work output in Eq.~\eqref{Waverage}, the standard deviation can be rewritten in terms of the local Hamiltonian and global ground state:
\be
\sigma^2=\braket{H_\Loc^2}_{\tilde0}-\braket{H_\Loc}_{\tilde0}^2.
\ee
We note that $\sigma$ is also the standard deviation for the quantum heat since $Q_l=W_l+\Delta$, where the local entanglement gap~$\Delta$ is a constant.

In what follows, we will apply the general formalism developed in this section, and summarized in Table~\ref{table}, to two types of interacting many-body systems: coupled qubits in Sec.~\ref{QBs} and coupled harmonic oscillators in Sec.~\ref{bosonsec}.

\begin{table}
\centering
\begin{tabular}{l c l}
Entangled ground state && $\ket{\tilde0}$\\
\hline
Local ground state && $\ket{0_\Loc}$\\
\hline
Local entanglement gap && $\Delta=E_{0_\Loc}-E_{\tilde0}$\\
\hline
Work output && $W=\braket{H_\Loc}_{\tilde0}-E_{0_\Loc}$\\
\hline
Quantum heat && $Q=W+\Delta=\braket{H_\Loc}_{\tilde0}-E_{\tilde0}$\\
\hline
Efficiency && $\eta=W/Q=W/(W+\Delta)$\\
\hline
Standard deviation && $\sigma=\sqrt{\braket{H_\Loc^2}_{\tilde0}-\braket{H_\Loc}_{\tilde0}^2}$
\end{tabular}
\caption{Summary of the most important results and notations for a generic many-body vacuum fluctuation engine.}
\label{table}
\end{table}

\section{Strongly interacting many-qubit engines}
\label{QBs}

\subsection{The two-qubit engine}
\label{2qb}

The minimal quantum system featuring an entangled ground state consists of two coupled qubits. As such, we consider two qubits, denoted by $\A$ and $\B$, coupled to one another via the horizontal components of their spins. Denoting by $\omega_j$ qubit~$j$'s transition frequency, and by $g$ the interqubit coupling strength, the local and interaction Hamiltonians read as (we set $\hbar=1$)
\begin{align}
&H_\Loc=\omega_\A\sigma_\A^+\sigma_\A^-+\omega_\B\sigma_\B^+\sigma_\B^-,\label{Hloc_2qb}\\
&H_\Int=\frac g2\sigma_\A^x\sigma_\B^x,\label{Hint_2qb}
\end{align}
where we have introduced the raising and lowering operators for qubit~$j$, $\sigma_j^+=\ket{1_j}\!\bra{0_j}$ and $\sigma_j^-=\ket{0_j}\!\bra{1_j}$, as well as its spin's $x$ component, $\sigma_j^x=\sigma_j^++\sigma_j^-$. We note that $\braket{0_j|\sigma_j^x|0_j}=\braket{1_j|\sigma_j^x|1_j}=0$, so the interaction terms can be turned off with no expected energy cost when the system is in a separable state.

As a result of the interaction between qubits~$\A$ and~$\B$, the eigenstates of the total Hamiltonian~$H=H_\Loc+H_\Int$ are entangled; we write them as
\begin{align}
&\ket{\phi^+}=\sin\phi\ket{00}+\cos\phi\ket{11},\\
&\ket{\psi^+}=\sin\psi\ket{01}+\cos\psi\ket{10},\\
&\ket{\psi^-}=\cos\psi\ket{01}-\sin\psi\ket{10},\\
&\ket{\phi^-}=\cos\phi\ket{00}-\sin\phi\ket{11},\label{phi-}
\end{align}
where the angles~$\phi$ and~$\psi$ satisfy
\begin{align}
&\tan\phi=\frac\gamma{1+\sqrt{1+\gamma^2}},\label{tanphi}\\
&\tan\psi=\frac\gamma{\delta+\sqrt{\delta^2+\gamma^2}},
\end{align}
with $\gamma=g/(\omega_\A+\omega_\B)$ and $\delta=(\omega_\A-\omega_\B)/(\omega_\A+\omega_\B)$, $0<\delta<1$. The corresponding eigenenergies are given by
\begin{align}
&E_\phi^\pm=\frac{\omega_\A+\omega_\B}2\left(1\pm\sqrt{1+\gamma^2}\right),\label{wphi}\\
&E_\psi^\pm=\frac{\omega_\A+\omega_\B}2\left(1\pm\sqrt{\delta^2+\gamma^2}\right)\label{wpsi},
\end{align}
with $E_\phi^-<E_\psi^-<E_\psi^+<E_\phi^+$. Using the notations in Table~\ref{table}, we write $\ket{\tilde0}=\ket{\phi^-}$ and $\ket{0_\Loc}=\ket{00}$. The definition of the local Hamiltonian in Eq.~\eqref{Hloc_2qb} sets the local ground-state energy at zero: $E_{0_\Loc}=0$; the local entanglement gap is then given by $\Delta=-E_{\tilde0}=-E_\phi^-$.

\begin{figure}
\centering
\includegraphics[width=\linewidth]{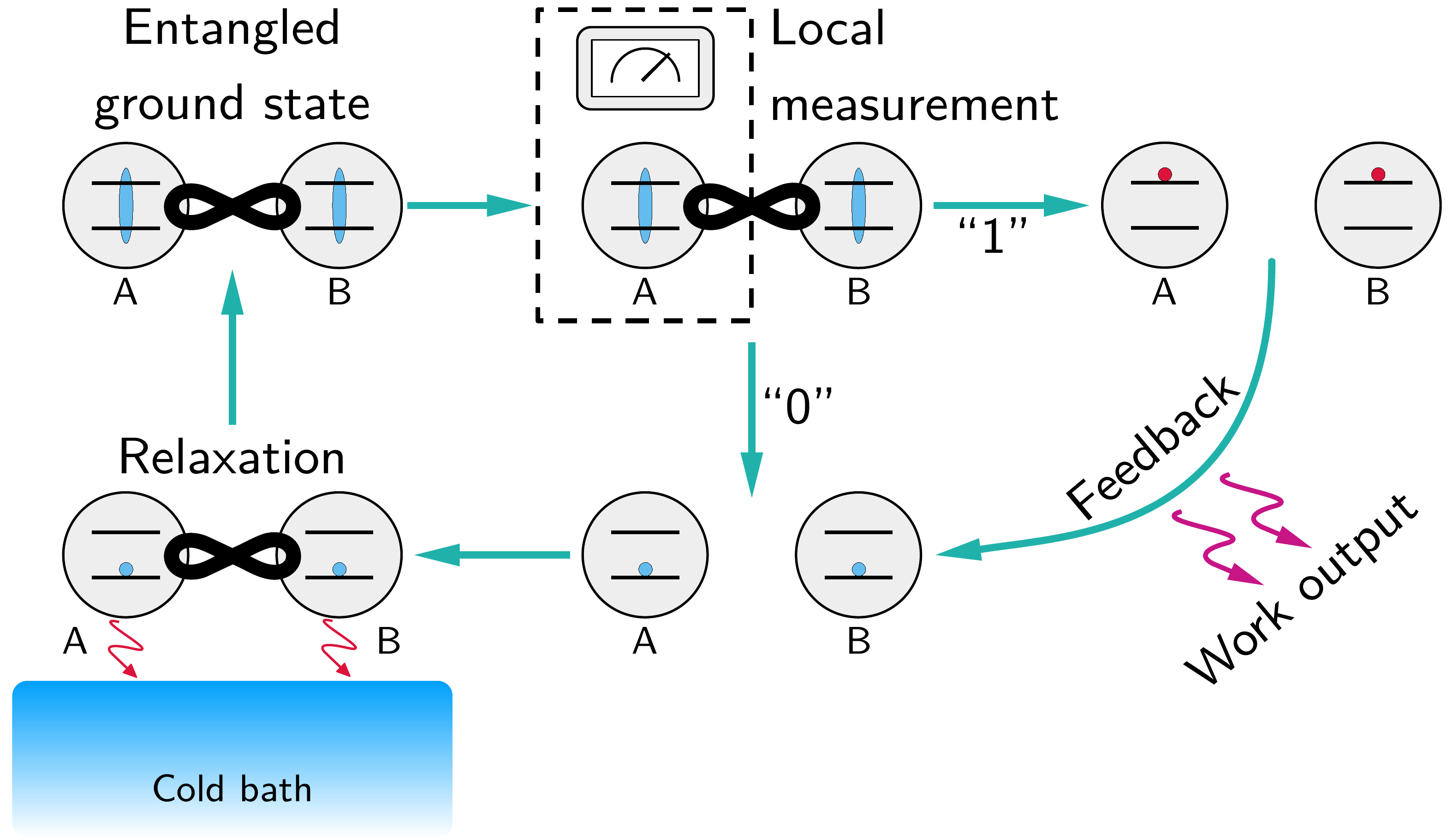}
\caption{The engine cycle for two coupled qubits: The two-qubit system is initially in its entangled ground state (top left). A local projective measurement is then performed on qubit~$\A$ (top middle). The joint state of the qubits after the measurement is either $\ket{11}$ (top right), or $\ket{00}$ (bottom middle). In the former case, work can be extracted by applying local pulses to each qubit, which takes the two-qubit system to state~$\ket{00}$ extracting work in the process. From there, it is put in contact with a cold bath so that it relaxes to its ground state (bottom left).}
\label{cycle_2qb}
\end{figure}

The engine cycle described in Sec.~\ref{General} applied to the two-qubit system considered here is depicted in Fig.~\ref{cycle_2qb}. One should note that a projective measurement on one the qubits, say $\A$, in the local eigenbasis~$\{\ket0,\ket1\}$ suffices to fully characterize the state of the two-qubit system: ``0'' corresponds to the two-qubit state~$\ket{00}$, and similarly ``1'' corresponds to $\ket{11}$. In the latter case, we turn off interactions with no energetic cost since $\braket{11|H_\Int|11}=0$, and we apply local pulses to each qubit so as to bring the two-qubit system to the local ground state~$\ket{00}$. In doing so, work is extracted as both qubits emit a stimulated photon at their resonant frequency. Using Eqs.~\eqref{phi-} and~\eqref{tanphi}, we find the probability for each outcome of the local measurement:
\begin{align}
&P_{00}=\cos^2\phi=\frac12\left(1+\frac1{\sqrt{1+\gamma^2}}\right),\\
&P_{11}=\sin^2\phi=\frac12\left(1-\frac1{\sqrt{1+\gamma^2}}\right).
\end{align}
Once the two-qubit system is in its local ground state, we turn interactions back on with no energetic cost given that $\braket{00|H_\Int|00}=0$, and we couple the system to a cold bath so that it relaxes to its entangled ground state. The cycle can then resume.  We note that to carry out the projective measurements in the local basis, an external system must couple strongly and quickly. This process is modeled in Appendix~\ref{2qb_dynamics} and we show meter coupling around a factor of $10$ larger than the system energies is sufficient to carry out an approximate projective measurement.

\begin{figure}
\centering
\includegraphics[width=\linewidth]{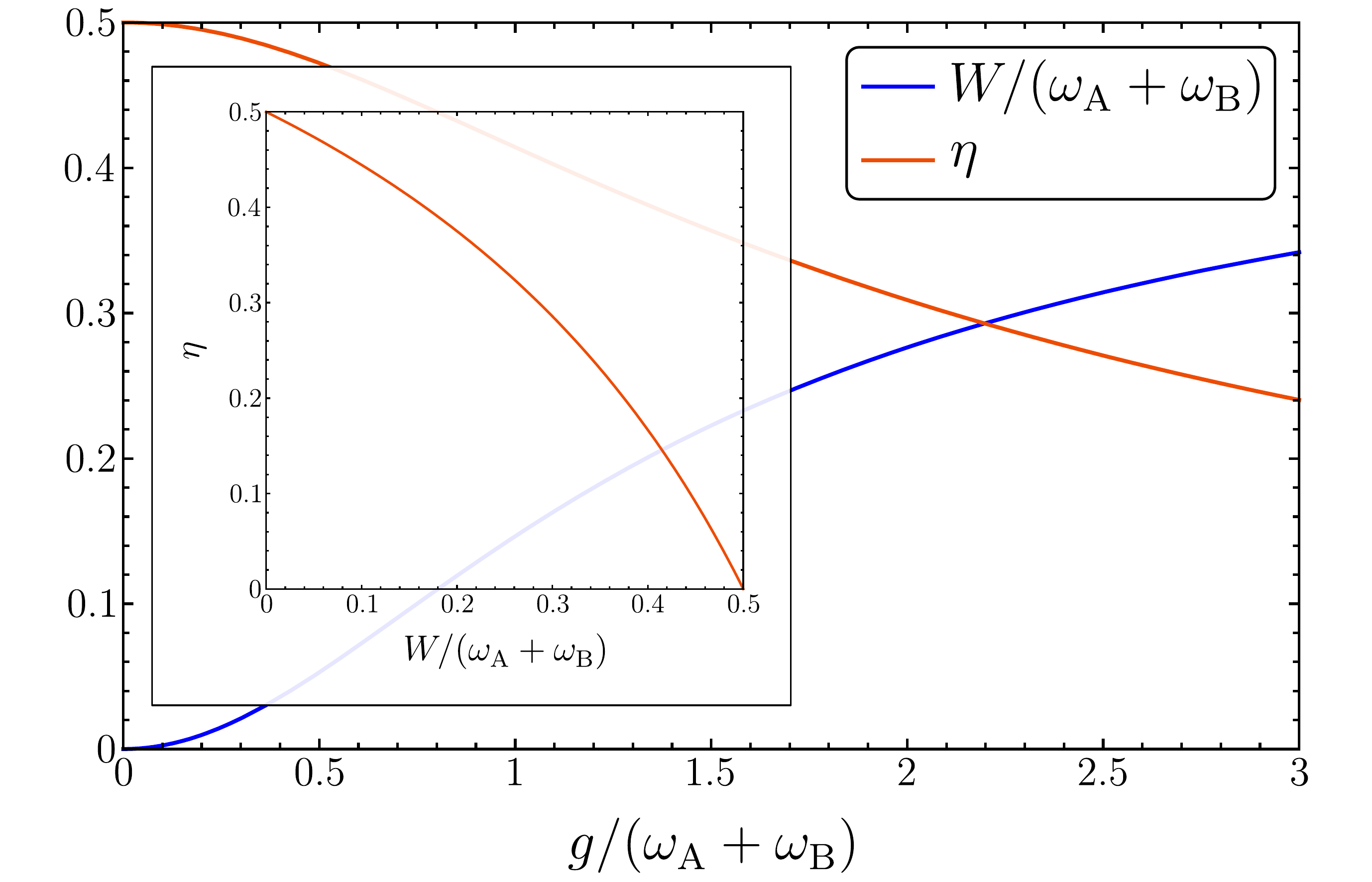}
\caption{Work ouput~(blue curve) and efficiency~(red curve) of the two-qubit engine as functions of the coupling between the two qubits. The inset shows the direct relation between work and efficiency.}
\label{Work_Eff}
\end{figure}

Using the general results in Table~\ref{table}, one can readily obtain the two-qubit engine work and efficiency. The average work output is given by
\be
W=\braket{H_\Loc}_{\tilde0}=\frac{\omega_\A+\omega_\B}2\left(1-\frac1{\sqrt{1+\gamma^2}}\right),
\label{work}
\ee
while the quantum heat necessary for the engine to operate reads as
\be
Q=W+\Delta=\frac{\omega_\A+\omega_\B}2\frac{\gamma^2}{\sqrt{1+\gamma^2}}.
\ee
This yields the efficiency:
\be
  \eta=\frac WQ=\frac1{1+\sqrt{1+\gamma^2}}.
\ee
The engine's work output and efficiency are plotted as functions of $\gamma$ in Fig.~\ref{Work_Eff}. In the weak coupling limit where $\gamma\ll1$, that is, $g\ll\omega_\A+\omega_\B$, the interaction Hamiltonian is negligible as compared with the local Hamiltonian. This means that the many-body ground state is almost equal to the local ground state: $\ket{\tilde0}\simeq\ket{0_\Loc}=\ket{00}$. As a consequence, the probability that the local measurement yields the excited state~$\ket{11}$ is vanishingly small, and almost every realization of the cycle results in no work extracted. The fact that the many-body and local ground states almost coincide also implies that the local entanglement gap closes: $\Delta\simeq0$. Coincidentally, the work output and local entanglement gap have the same asymptotic behavior as $\gamma$ vanishes. The efficiency then tends to $1/2$---its maximum value---in the weak coupling limit.

As $\gamma$ increases, the work output increases and the efficiency decreases. The trade-off between these quantities is described through the relation
\be
\eta=1-\frac{\omega_\A+\omega_\B}{2(\omega_\A+\omega_\B-W)},
\ee
which is plotted in the inset of Fig.~\ref{Work_Eff}.

In the deep strong coupling limit where $\gamma\gg1$, or equivalently $g\gg\omega_\A+\omega_\B$, the interacting ground state becomes a Bell state: $\ket{\tilde0}\simeq(\ket{00}-\ket{11})/\sqrt2$. As such, the local ground state~$\ket{00}$ and the excited state~$\ket{11}$ are equiprobable outcomes of the local measurement. Work extraction then occurs for half of the cycle's realizations, so the average work output is $(\omega_\A+\omega_\B)/2$. This is the upper bound for the engine's average work output. Conversely, since the local entanglement gap grows linearly with $\gamma$ in the deep strong coupling limit, the efficiency vanishes.

Finally, fluctuations of the work output between different realizations of the cycle are described through the standard deviation~$\sigma$. We obtain (see Table~\ref{table})
\be
\sigma=\sqrt{\braket{H_\Loc^2}_{\tilde0}-\braket{H_\Loc}_{\tilde0}^2}=\frac{\omega_\A+\omega_\B}2\frac\gamma{\sqrt{1+\gamma^2}}.
\ee
In the weak coupling regime ($\gamma\ll1$), the standard deviation scales linearly with $\gamma$ while the average extracted work is quadratic, so fluctuations swamp the average in this limit. In contrast, both the work output's average and standard deviation saturate to the value~$(\omega_\A+\omega_\B)/2$ in the deep strong coupling regime ($\gamma\gg1$).

To compute the power generated through a realization of the cycle, it is necessary to model its dynamics, which is discussed in Appendix~\ref{2qb_dynamics}.

\subsection{The qubit chain engine}
\label{Nqb}

We extend the working principle of the two-qubit engine to $N$ coupled qubits. We will focus here on the case of a closed chain with periodic boundary conditions (Appendix~\ref{OpenChain} provides some details about the case of an open chain). We then consider a closed chain of $N$ qubits with transition frequencies~$\omega$ coupled to their nearest neighbors with the coupling amplitude~$g$. The local and interaction Hamiltonians are then written as
\begin{align}
&H_\Loc=\omega\sum_{j=1}^N\sigma_j^+\sigma_j^-,\label{Hloc_chain}\\
&H_\Int=\frac g2\sum_{j=1}^N\sigma_j^x\sigma_{j+1}^x\label{Hint_chain},
\end{align}
with $\sigma_{N+1}^x=\sigma_1^x$. The procedure considered previously to extract work using local operations can be applied here as well: The chain, initially in its entangled ground state, is projected onto the local eigenbasis. To do so, local measurements must be performed on all but one of the qubits within the chain as can be seen in Eq.~\eqref{Zstate} below. The interaction Hamiltonian is then turned off at no energetic cost. Each qubit found in an excited state is flipped using a local pulse, which transfers energy to the field driving the pulse. With the chain now in its local ground state, interactions are turned back on at no energetic cost, and the system is coupled to a cold environment which relaxes it to its many-body ground state. The cycle can then restart.

The total Hamiltonian~$H=H_\Loc+H_\Int$ is analogous to the transverse-field Ising model, where $\omega$ and $g$ would respectively represent the intensity of the transverse magnetic field and the exchange parameter characterizing the interaction between neighboring spins. This connection to quantum magnetism provides helpful tools to assess the qubit chain engine's performance. The work output is given by the expectation value of the local Hamiltonian in the ground state (see Table~\ref{table}). In the Ising model language, this corresponds to the energy from the transverse magnetic field which is proportional to the transverse magnetization. To further obtain the engine's efficiency, it is necessary to calculate the local entanglement gap  which is readily deduced from the ground-state energy (see Table~\ref{table}). One can then fully characterize the performance of a many-qubit engine from the transverse magnetization and ground-state energy in the corresponding Ising model.

\begin{figure}
\includegraphics[width=\linewidth]{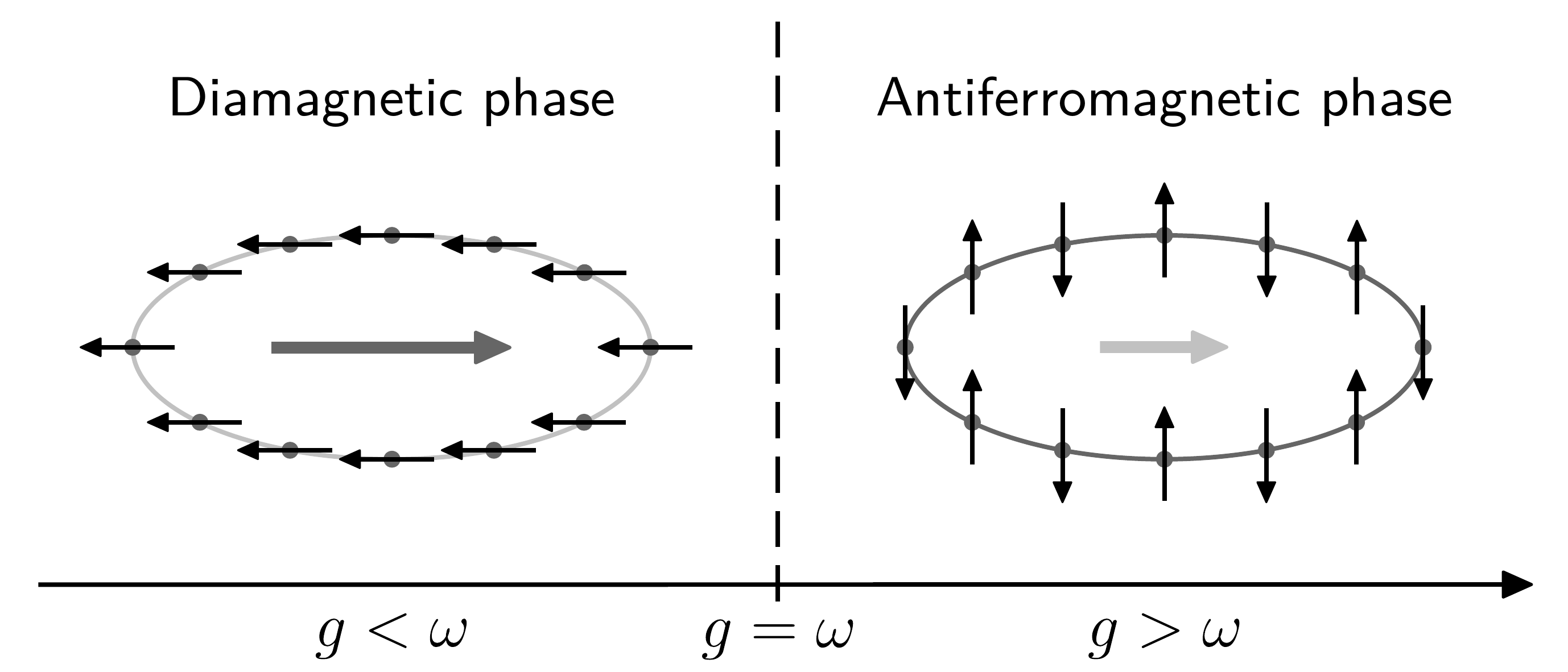}
\caption{Schematic representation of the quantum phase transition for the transverse-field Ising model in one dimension. When $g<\omega$ (left), the local Hamiltonian is dominant, and all individual spins point in the direction opposite to that of the transverse field (diamagnetic phase). Conversely, when $g>\omega$ (right), the interaction Hamiltonian becomes the dominant contribution. As a result, spins tend to antialign with their neighbors (antiferromagnetic phase).}
\label{phases}
\end{figure}

Furthermore, the one-dimensional transverse-field Ising model considered here undergoes a quantum phase transition~\cite{lieb1961,suzuki2012,he2017,sachdev2011}. With the conventions used to define the Hamiltonian in Eqs.~\eqref{Hloc_chain} and~\eqref{Hint_chain}, this is a transition from a diamagnetic phase~($g<\omega$) to an antiferromagnetic phase~($g>\omega$) as depicted in Fig.~\ref{phases}. The magnetic order in both of these phases correspond to different behaviors of the longitudinal magnetization, but other quantities also bear the signature of the phase transition. This includes the transverse magnetization and ground-state energy from which the engine's work output and efficiency are inferred.

An analytical solution for the one-dimensional transverse-field Ising model can be obtained using a Jordan--Wigner transformation which maps spins to fermions~\cite{lieb1961,suzuki2012,he2017,sachdev2011}. We then introduce the fermionic operators
\be
  c_j=\exp\left(\ii\pi\sum_{k=1}^{j-1}\sigma_k^+\sigma_k^-\right)\sigma_j^-,
  \label{JordanWigner}
\ee
where the first term to the right-hand side of the equation above ensures that fermionic anticommutation relations are satisfied. We next move to momentum space~\cite{suzuki2012,he2017,sachdev2011}:
\be
  c_p=\frac1{\sqrt N}\sum_{j=1}^Nc_j\e^{-j\ii p},
\ee
and obtain
\be
  H=\sum_p\Big(
\begin{aligned}[t]
  &(\omega+g\cos p)\big(c_p^\dagger c_p-c_{-p}c_{-p}^\dagger\big)\\
  &+\ii g\sin p\,\big(c_p^\dagger c_{-p}^\dagger-c_{-p}c_p\big)\Big)+\frac{N\omega}2.
\end{aligned}
\ee
The Hamiltonian is then diagonalized by performing the adequate Bogoliubov transformation: $\zeta_p=u_pc_p+\ii v_pc_{-p}^\dagger$, with\footnote{If $p$ is a multiple of $\pi$, the Bogoliubov transformation is unnecessary; we then have $u_p=1$, $v_p=0$, and $\Omega_p=\omega+g\cos p$.}
\begin{align}
  &u_p=\frac{g\abs{\sin p}}{\sqrt{2\Omega_p(\Omega_p-\omega-g\cos p)}},\\
  &v_p=\frac{\sgn p}{\sqrt2}\sqrt{1-\frac{\omega+g\cos p}{\Omega_p}},\\
  &\Omega_p=\sqrt{\omega^2+g^2+2\omega g\cos p}.
\end{align}
The fact that $u_p^2+v_p^2=1$ ensures that fermionic commutation relations are satisfied for the quasiparticle operators: $\{\zeta_p,\zeta_q^\dagger\}=\delta_{pq}$, $\{\zeta_p, \zeta_q\}=0$. The transverse-field Ising Hamiltonian can then be rewritten as a free fermion Hamiltonian:
\be
  H=\sum_p\Omega_p\left(\zeta_p^\dagger\zeta_p-\frac12\right)+\frac{N\omega}2.
  \label{Hzeta}
\ee
It must be noted that the set of possible momentum values in the equation above depend on the number of excitations~\cite{he2017}, or, equivalently, the number of Jordan--Wigner fermions, $n=\sum_j\sigma_j^+\sigma_j^-=\sum_jc_j^\dagger c_j$. Namely, $p=(2m-1)\pi/N$ for an even number of excitations, while $p=2m\pi/N$ for an odd number of excitations, where $m$ is an integer ranging from $-\lfloor(N-1)/2\rfloor$ to $\lfloor N/2\rfloor$. More details about the derivation of Eq.~\eqref{Hzeta} are given in Appendix~\ref{Details}.

The ground state of $H$ is the quasiparticle vacuum, that is, the state~$\ket{\tilde0}$ such that $\zeta_p\ket{\tilde0}=0$ for all $p$. We find that it can be written as
\be
\ket{\tilde0}=\prod_p\left(\sqrt{u_p}-\frac{\ii v_p}{\sqrt{u_p}}c_p^\dagger c_{-p}^\dagger\right)\ket{0_\Loc}
\label{Zstate}
\ee
where the local ground state~$\ket{0_\Loc}=\otimes_j\ket0$ corresponds to the vacuum state for Jordan--Wigner fermions, that is, the state where all qubits are in their individual ground state. The entangled ground state~$\ket{\tilde0}$ in Eq.~\eqref{Zstate} above is a linear combination of states with an even number of excitations. As a consequence, the state of the $N$-qubit chain in the local basis can be deduced from local measurements on $N-1$ of the qubits within the chain. This is similar to the two-qubit case where it was sufficient to measure one of the qubits as depicted in Fig.~\ref{cycle_2qb}.

One can then read the ground-state energy~$E_{\tilde0}$ from Eq.~\eqref{Hzeta}. Since the local Hamiltonian in Eq.~\eqref{Hloc_chain} has been defined such that its ground-state energy~$E_{0_\Loc}$ is set at zero, we find that the local entanglement gap is given by (see Table~\ref{table})
\be
\Delta=-E_{\tilde0}=\frac12\left(\sum_p\Omega_p-N\omega\right).
\label{Delta_chain}
\ee
As the entangled ground state~$\ket{\tilde0}$ in Eq.~\eqref{Zstate} is a linear combination of states with an even number of excitations, the values of the momentum~$p$ in the summation of Eq.~\eqref{Delta_chain} above are such that: $p=(2m-1)\pi/N$, with $m$ an integer between $-\lfloor(N-1)/2\rfloor$ and $\lfloor N/2\rfloor$.

We now calculate the engine's work output which, as shown in Table~\ref{table}, is equal to the expectation value for the local Hamiltonian in the interacting ground state. We then rewrite the local Hamiltonian in terms of quasiparticle operators:
\be
  H_\Loc=\omega n=\omega\sum_p\left(u_p\zeta_p^\dagger+\ii v_p\zeta_{-p}\right)\big(u_p\zeta_p-\ii v_p\zeta_{-p}^\dagger\big),
  \label{Hloc_qp}
\ee
where we have expressed the number operator in momentum space: $n=\sum_j\sigma_j^+\sigma_j^-=\sum_pc_p^\dagger c_p$, and inverted the Bogoliubov transformation. We then obtain
\be
  W=\braket{H_\Loc}_{\tilde0}=\omega\sum_pv_p^2=\frac\omega2\Bigg(N-\sum_p\frac{\omega+g\cos p}{\Omega_p}\Bigg).
  \label{W_chain}
\ee
The engine's efficiency can now be deduced from the work output and the local entanglement gap using the relation~$\eta=W/(W+\Delta)$ from Table~\ref{table}.

Analytical results can be obtained in the thermodynamic limit where $N\to\infty$: The summations in Eqs.~\eqref{Delta_chain} and~\eqref{W_chain} become elliptic integrals, and we find
\begin{widetext}
\begin{align}
  &\frac\Delta N\underset{N\to\infty}\simeq\frac{\abs{\omega-g}}\pi E\left(-\frac{4\omega g}{(\omega-g)^2}\right)-\frac\omega2,\label{Delta_inf}\\
  &\frac WN\underset{N\to\infty}\simeq\frac\omega2-\frac{\sgn(\omega-g)}{2\pi}\left((\omega-g)E\left(-\frac{4\omega g}{(\omega-g)^2}\right)+(\omega+g)K\left(-\frac{4\omega g}{(\omega-g)^2}\right)\right),\label{W_inf}
\end{align}
\end{widetext}
where $K(x)$ and $E(x)$ are the complete elliptic integrals first and second kinds:
\begin{align}
  &K(x)=\int_0^{\frac\pi2}\frac{\mathrm d\theta}{\sqrt{1-x^2\sin^2\theta}},\\
  &E(x)=\int_0^{\frac\pi2}\mathrm d\theta\,\sqrt{1-x^2\sin^2\theta}.
\end{align}
Both $\Delta$ and $W$ above are nonanalytic at $g=\omega$: the second derivative of $\Delta$ and the first derivative of $W$ diverge at this point. As a consequence, the engine's work output and efficiency both exhibit a vertical tangent at the critical point~$g=\omega$. Nevertheless, these quantities have well-defined critical values:
\begin{align}
  &\frac W{N\omega}\underset{g=\omega}=\frac12-\frac1\pi\approx0.182,\\
  &\eta\underset{g=\omega}=\frac\pi2-1\approx0.571.
\end{align}

One can gain insights about the scaling of work and efficiency with the number of qubits by analyzing the limiting cases~$g\ll\omega$ (weak coupling limit, diamagnetic phase) and~$g\gg\omega$ (deep strong coupling limit, antiferromagnetic phase). In the weak coupling limit~$g\ll\omega$, we find that the local entanglement gap and the work output behave in a similar way, scaling linearly with the number of qubits. Indeed, for $N>2$, we have\footnote{We refer the reader to Sec.~\ref{2qb} for the two-qubit case.}
\be
  \Delta\simeq W\simeq\frac{Ng^2}{8\omega}.
\ee
As a consequence, the efficiency is independent of the number of qubits and is simply given by: $\eta\simeq1/2$.

In the deep strong coupling limit~$g\gg\omega$, we find that the local entanglement gap and work output again scale linearly with the number of qubits:
\begin{align}
  &\Delta\simeq\left(2\left\lfloor\frac N2\right\rfloor-\frac N2\right)g=
  \begin{cases}
  Ng/2,&\text{if $N$ is even},\\
  (N/2-1)g,&\text{if $N$ is odd},
  \end{cases}\label{Delta_DS}\\
  &W\simeq\left(2\left\lfloor\frac N2\right\rfloor-\frac N2\right)\omega=
  \begin{cases}
  N\omega/2,&\text{if $N$ is even},\\
  (N/2-1)\omega,&\text{if $N$ is odd}.
  \end{cases}\label{W_DS}
\end{align}
Again, the efficiency is independent on the number of qubits here: $\eta\simeq\omega/g$. The discrepancy observed between even and odd values of $N$ above is a consequence of geometrical frustration in the antiferromagnetic phase: When $g\gg\omega$, the interaction Hamiltonian dominates which favors the antialignment of neighboring spins. For an even number of spins, it is possible to find a configuration where each spin is antialigned with its neighbors, while such a configuration does not exist for an odd number of spins. This results in different structures for the antiferromagnetic ground state in each case~\cite{he2017}. This difference manisfests in the quasiparticle picture of Eq.~\eqref{Hzeta} through the fact that one of the quasiparticle energies, namely, $\Omega_\pi=\omega-g$, is negative for $g>\omega$ when $N$ is odd. In contrast, all quasiparticle energies remain positive when $N$ is even; this is because $p=\pi$ is not a relevant value for the momentum in this case.

\begin{figure}
\centering
\includegraphics[width=\linewidth]{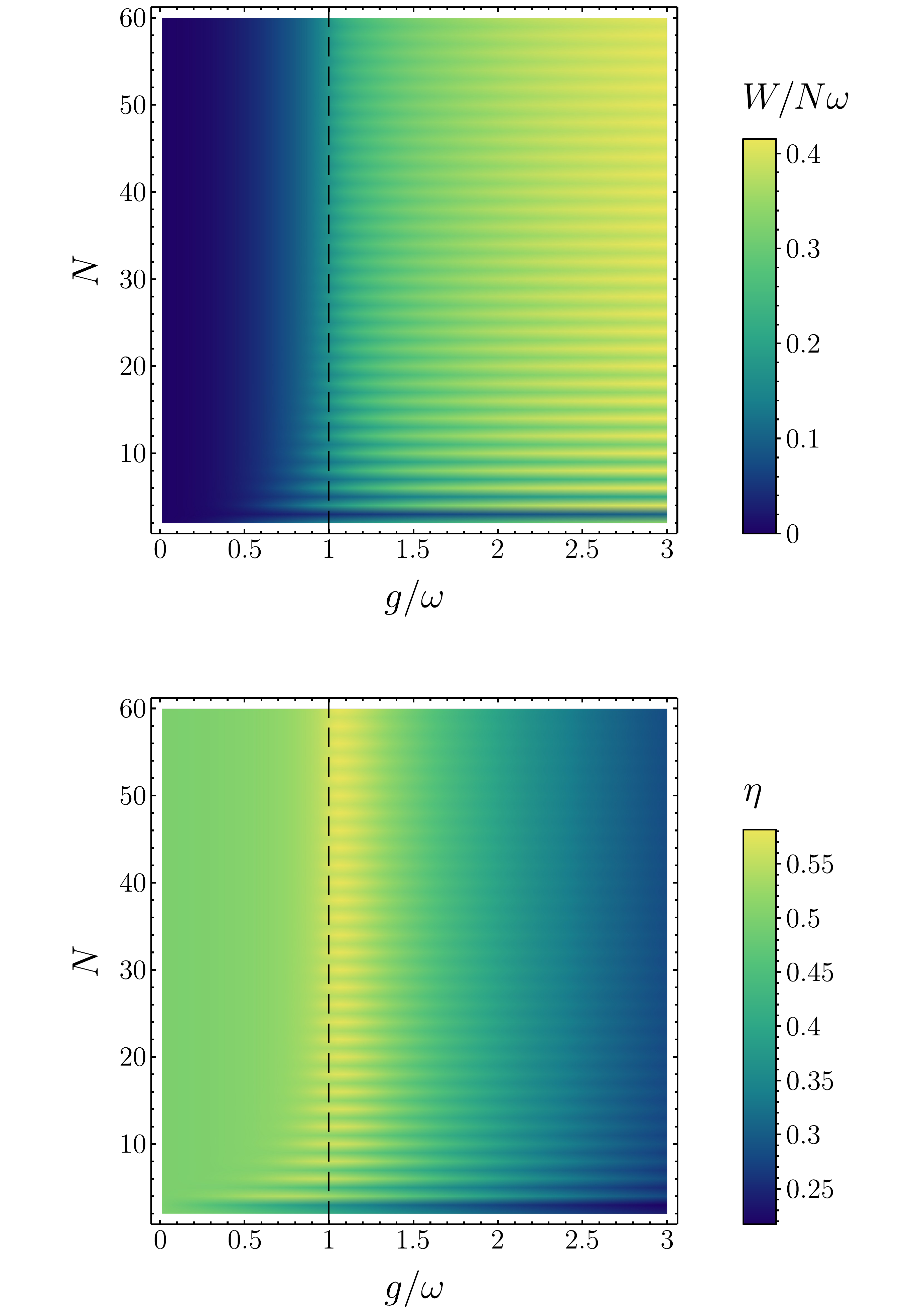}
\caption{Density plots of the engine's work output per qubit~(upper panel) and efficiency~(lower panel) as functions of the coupling strength~$g$ and the number of qubits~$N$. Both plots confirm that the scalings observed in the weak coupling and deep strong coupling limits. The horizontal stripe pattern appearing on these density plots shows the influence of the parity of $N$ on the results, which tends to vanish as $N$ increases. The quantum phase transition at $g=\omega$ (black dashed line) is clearly visible on both plots with a sharp increase of the work output and a peak of the efficiency.}
\label{W-eff_chain}
\end{figure}

As shown in Fig.~\ref{W-eff_chain}, the observations made above concerning the engine's work output and efficiency still hold in the general case: The work output scales linearly with the number of qubits, while the efficiency is essentially constant. One should also note that both density plots in Fig.~\ref{W-eff_chain} exhibit a horizontal stripe pattern which corresponds to the dependence of work and efficiency on the parity of the number of qubits within the chain. More precisely, the different results obtained for an even or odd number of qubits are a consequence of geometric frustration in the antiferromagnetic phase as explained previously. According to Eqs.~\eqref{Delta_DS} and~\eqref{W_DS}, such a difference is of order~$1/N$.

Fig.~\ref{W-eff_chain} also highlights the impact of the quantum phase transition occurring at $g=\omega$ on the engine's performance. One can clearly distinguish two regimes for the work output~(upper panel): The work output vanishes for $g=0$ and remains small in the diamagnetic phase~($g<\omega$). It then abruptly increases around the critical point, with a vertical tangent at $g=\omega$. In the antiferromagnetic phase~($g>\omega$), the work output plateaus as it approaches its maximum value given in Eq.~\eqref{W_DS}. As for the efficiency, it is essentially constant in the diamagnetic phase~($g<\omega$) with $\eta\approx1/2$. Similar to the work output, the quantum phase transition is marked by an abrupt increase of the efficiency, with a vertical tangent at $g=\omega$. The efficiency peaks for $g\gtrsim\omega$, and then decreases in the antiferromagnetic phase~($g>\omega$) to eventually vanish in the deep strong coupling limit. We note that the observation of a clear trade-off between work and efficiency made in the case of two coupled qubits (see Fig.~\ref{Work_Eff}) cannot be generalized to a longer chain: For a larger number of coupled qubits, maximum efficiency is reached at nonzero coupling. Interestingly, we find that the work output at maximum efficiency is not zero in this case, and it is in fact relatively close to its maximum value. This is a consequence of the sharp increase of the efficiency around $g=\omega$ caused by the divergence of its first derivative at the critical point. Our results corroborate recent analyses which showed that the thermodynamic performance of various quantum machines operating close to a quantum critical point bears the signature of the corresponding phase transition~\cite{Campisi2016,Ma2017,Purkait2022,Fadaie2018,Piccitto2022}. In particular, an enhancement of efficiency around the critical point has been noted in Refs.~\cite{Campisi2016,Ma2017,Purkait2022}.

We conclude this section by addressing the question of fluctuations. The work output standard deviation~$\sigma$ is calculated using the relation in Table~\ref{table} and the expression for the local Hamiltonian in Eq~\eqref{Hloc_qp}:
\be
\sigma^2=2\omega^2\sum_pu_p^2v_p^2=\frac{\omega^2g^2}2\sum_p\frac{\sin^2p}{\Omega_p^2}.
\label{sigma_chain}
\ee
Once again, we scrutinize the weak and deep strong coupling limits to obtain insights about the scaling of the standard deviation with the number of qubits within the chain. We find
\begin{align}
&\sigma\underset{g\ll\omega}\simeq\frac{\sqrt N}2g,\label{sigma_weak}\\
&\sigma\underset{g\gg\omega}\simeq\frac{\sqrt N}2\omega,\label{sigma_strong}
\end{align}
which suggests that $\sigma$ scales like $\sqrt N$. In fact, the results in Eqs.~\eqref{sigma_weak} and~\eqref{sigma_strong} above accurately approximate the standard deviation within the whole diamagnetic ($g<\omega$) and antiferromagnetic ($g>\omega$) phase respectively. In the thermodynamic limit~$N\to\infty$, the approximation matches the exact result which is obtained by replacing the summation in Eq.~\eqref{sigma_chain} by an integral:
\be
\frac\sigma{\sqrt N}\underset{N\to\infty}\simeq\frac12\min(\omega,g).
\ee
We note that the only effect of the phase transition on the standard deviation is a slope discontinuity at the critical point. Further, we observe that $\sigma$ scales as $\sqrt N$ which shows that the fluctuations of the work output become negligible in comparison to its average value for $N\to\infty$.

\section{The bosonic vacuum engine}
\label{bosonsec}

\subsection{The two-oscillator engine}
\label{2osc}

In the previous section, we showed the physics of the qubit chain could be mapped to free fermions and solved. Here, we consider the bosonic case, and we begin with the simplest version of two coupled harmonic oscillators:
\be
H=\frac12\left(p_1^2+p_2^2\right)+\frac{k_0}2\left(x_1^2+x_2^2\right)+\frac g2\left(x_1-x_2\right)^2,
\label{ham1}
\ee
where $x_j$ and $p_j$ respectively denote the position and momentum for oscillator~$j$, $k_0$ is the force constant for both oscillators, and $g$ is the coupling constant. The Hamiltonian in Eq.~\eqref{ham1} above can be written as $H=H_\Loc+H_\Int$, where the local and interaction Hamiltonians are given by
\begin{align}
&H_\Loc=\frac12\left(p_1^2+p_2^2\right)+\frac12(k_0+g)\left(x_1^2+x_2^2\right),\\
&H_\Int=-gx_1x_2.
\end{align}
We denote by $\ket{n_1,n_2}$ the local eigenstates, where $n_j$ is the energy quantum number for local oscillator~$j$. Obviously, the local ground state is $\ket{0_\Loc}=\ket{0,0}$, and its energy is given by the local oscillator frequency: $E_{0_\Loc}=\omega=\sqrt{k_0+g}$. Furthermore, we note that the local eigenstates satisfy the condition~$\braket{n_1,n_2|H_\Int|n_1,n_2}=0$ because of parity, indicating that there is no expected energy in the coupling term.

Introducing the sum and difference variables, $x_\pm = (x_1 \pm x_2)/\sqrt{2}$, and quantizing the total Hamiltonian~$H=H_\Loc+H_\Int$ in the standard fashion, it takes the form of two independent oscillators:
\be
H=-\frac12\left(\frac{\partial^2}{\partial x_+^2}+\frac{\partial^2}{\partial x_-^2}\right)+\frac12\omega_+^2x_+^2 + \frac12\omega_-^2x_-^2,
\ee
with the natural frequencies~$\omega_+=\sqrt{k_0}$ and~$\omega_-=\sqrt{k_0+2g}$. In this new basis, it is clear that the ground-state energy is $E_{\tilde0}=(\omega_++\omega_-)/2$. The associated ground-state wave function is given by
\be
\psi_{\tilde 0}(x_+, x_-) = \left(\frac{\omega_+ \omega_-}{\pi^2}\right)^{1/4}\e^{-(\omega_+ x_+^2+\omega_- x_-^2)/2}.
\ee

Following the engine cycle considered in previous sections, we now make local energy measurements of each oscillator, projecting each of them onto local energy eigenstates with energies $E_{n_1,n_2} =(n_1+n_2+ 1)\omega$. If oscillator~$j$ is found in an excited state~$\ket{n_j}$ with $n_j\ne0$, the excess energy can be transferred to a battery with local operations which lower the oscillator to its ground state. Once in the local ground state $\ket{0_\Loc} = \ket{0,0}$, the interaction with the cold environment relaxes the coupled oscillators to the entangled ground state~$\ket{\tilde0}$, and the cycle resets. 

The engine's work and efficiency can be obtained using the relations in Table~\ref{table}. The average work output of the engine corresponds to the energy transferred from the local oscillators to the idealized energy source during local operations. It is given by
\be
W=\braket{H_\Loc}_{\tilde0}-E_{0_\Loc}.
\ee 
The Gaussian nature of the integrals permits an exact calculation of the expectation value:
\be
\braket{H_\Loc}_{\tilde0}=\frac{\omega_+ + \omega_-}{4}\left(\frac{\omega^2}{\omega_+\omega_-}+1\right).
\label{Eloc2}
\ee
Thus, the total work is
\be
W=\frac{\omega_++\omega_-}{4}\left(\frac{\omega^2}{\omega_+\omega_-}+1\right)-\omega.
\ee
The amount of quantum heat given to the system by the measurement is the difference of the energy before and after the measurement, given, on average, by
\be
Q=\braket{H_\Loc}_{\tilde0}-E_{\tilde0}= \frac{\omega_+ + \omega_-}{4}\left(\frac{\omega^2}{\omega_+\omega_-}-1\right).
\ee
The efficiency of the engine, defined as $\eta=W/Q$, is controlled by the local entanglement gap:
\be
\Delta=W-Q=E_{0_\Loc}-E_{\tilde0}.
\ee
As for fluctuations, we calculate the work output standard deviation~$\sigma$, and we find (see Table~\ref{table})
\be
\sigma=\sqrt{\braket{H_\Loc^2}_{\tilde0}-\braket{H_\Loc}_{\tilde0}^2}=\frac{\omega g}{2\omega_+\omega_-}.
\label{sigma-2osc}
\ee

\begin{figure}
\centering
\includegraphics[width=\linewidth]{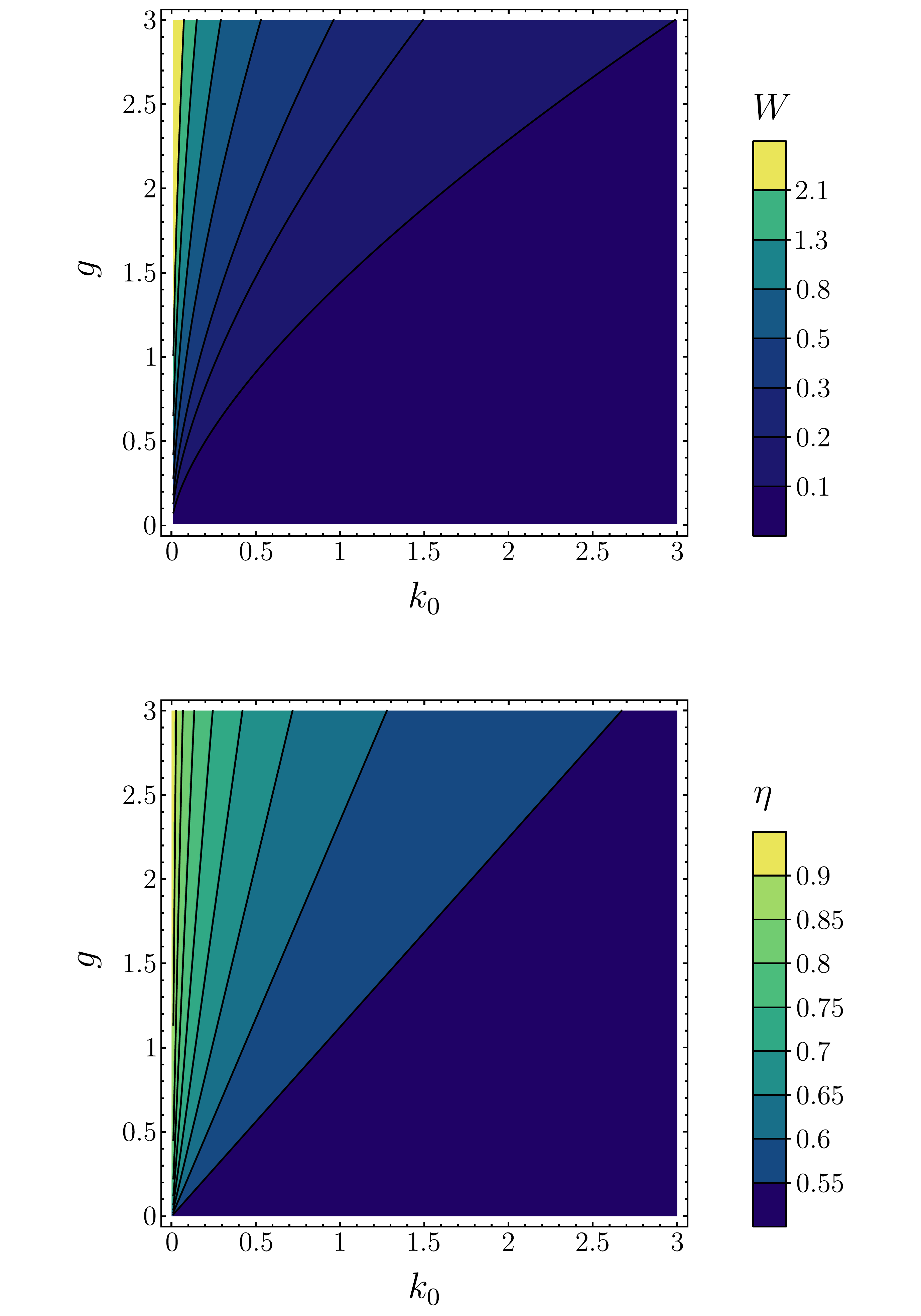}
\caption{Contour plots of the work output (upper panel) and efficiency (lower panel) of the two-oscillator engine versus $k_0$ and $g$.}
\label{fig:2osc}
\end{figure}

The work output and efficiency are plotted in contour plots for varying $k_0$ and $g$ in Fig.~\ref{fig:2osc}. Interestingly, both quantities increase as $k_0\to0$ and $g\to\infty$. In this limit, $W\simeq g/4\sqrt{k_0}$ while $\eta\simeq1-(4-2\sqrt{2}) \sqrt{k_0/g}$. We see that a perfect efficiency engine is possible in this situation with the caveat that the standard deviation diverges since $\sigma\simeq g/2\sqrt{2k_0}$.

It is also of interest to find the explicit probabilities of finding the each excitation~$\ket{n_1,n_2}$ as the result of a local measurement on the system in its entangled ground state:
\be
P_{n_1,n_2}=\abs{\braket{n_1,n_2|\tilde0}}^2.
\ee

We consider the state overlap
\be
\braket{n_1,n_2|\tilde0}=
\begin{aligned}[t]
&\sqrt{\frac{\omega}{2^{n_1+n_2}\pi\,n_1!\,n_2!}}\int\mathrm dx_1 \mathrm dx_2\,\e^{-\omega (x_1^2+x_2^2)/2}\\
&\times H_{n_1}(\sqrt{\omega} x_1)H_{n_2}(\sqrt{\omega} x_2)\psi_{\tilde0}(x_1, x_2),
\end{aligned}
\ee
where $H_n(z)$ are the Hermite polynomials. These state overlaps can be found with a joint generating function,
\be
Z(t_1,t_2)=\sum_{n_1,n_2=0}^\infty\frac{t_1^{n_1}t_2^{n_2}}{\sqrt{n_1!\,n_2!}} \braket{n_1,n_2|\tilde0}.
\ee
Using the identity $\sum_{n=0}^\infty H_n(x) t^n/n! = \e^{2xt-t^2}$, we find the result
\be
Z(t_1,t_2)=\frac{2\sqrt{\omega}(\omega_+ \omega_-)^{1/4}\e^{a(t_1^2+t_2^2)/2+bt_1t_2}}{\sqrt{(\omega+\omega_+)(\omega+\omega_-)}},
\ee
where
\begin{align}
&a=\frac{\omega(2 \omega+\omega_++\omega_-)}{(\omega+\omega_+)(\omega+\omega_-)},\\
&b=\frac{\omega (\omega_--\omega_+)}{(\omega+\omega_+)(\omega+\omega_-)}.
\end{align}

\subsection{The coupled oscillator network engine}
\label{Nosc}

We now generalize the previous treatment to an array of $N$ oscillators, linearly coupled to any other oscillator in the array. We follow the analysis of a related problem in Ref.~\cite{srednicki1993entropy}. The system is described by the Hamiltonian,
\be
H = \frac12\sum_jp_j^2+\frac12\sum_{j,k}x_jK_{jk}x_k,
\label{H_array}
\ee
where the sums all range from $1$ to $N$, and $K$ is a real symmetric matrix with positive eigenvalues. The Hamiltonian in Eq.~\eqref{H_array} above is broken into a local and interacting parts as follows
\begin{align}
&H_\Loc=\frac12\sum_j\left(p_j^2+K_{jj}x_j^2\right),\\
&H_\Int=\frac12\sum_{j \ne k}x_jK_{jk}x_k.\label{Hint_array}
\end{align}
The matrix $K$ can be diagonalized by an orthogonal matrix $O$ as $K=O^\mathrm TK_\mathrm DO$, and the diagonal matrix $K_\mathrm D$ has positive eigenvalues $\{k_m\}$, so $\Omega_m= \sqrt{k_m}$ is the natural frequency of normal mode~$m$. The normalized many-body ground-state wave function is then given by
\be
\psi_{\tilde0}(x)=\left(\frac{\det\Omega}{\pi^N}\right)^{1/4}\e^{-x\cdot\Omega\cdot x/2},
\ee
where $\Omega=O^\mathrm TK_\mathrm D^{1/2}O$ is the square root of $K$, and the vector $x$ has components $x_j$. This is generally an entangled state. 

The engine protocol described in the previous sections can be generalized here: Local energy measurements are performed on each local oscillator, with resulting energy
\be
E_{n_1,\dots,n_N}=\sum_j\left(n_j+\frac12\right)\sqrt{K_{jj}},
\ee
where $\sqrt{K_{jj}}$ is the natural frequency of local oscillator~$j$. The globally entangled nature of the ground state permits the possibility of finding the local oscillators in locally excited states~\cite{jordan2004}. The interaction Hamiltonian in Eq.~\eqref{Hint_array} is then turned off with no energetic cost. Any excess energy for a local oscillator found in an excited state is transferred to a battery with local operations, extracting its energy and lowering the oscillator to its ground state. Once the oscillator array is in its local ground state~$\ket{0_\Loc}=\ket{n_1=0,\dots,n_N=0}$, interactions are turned back on with no energetic cost. The system is finally let to relax to its many-body ground state, closing the cycle.

As shown in Table~\ref{table}, the extracted work is given by
\be
W=\braket{H_\Loc}_{\tilde0}-E_{0_\Loc},
\label{wosc}
\ee
where the local ground-state energy is
\be
E_{0_\Loc}=\frac12\sum_j\sqrt{K_{jj}}. 
\ee
The expectation value in Eq.~\eqref{wosc} may be calculated using multidimensional Gaussian integration to find
\be
\braket{H_\Loc}_{\tilde0}=\frac14\sum_j (K_{jj}(\Omega^{-1})_{jj}+\Omega_{jj}),
\label{mainresult}
\ee
which is our main result. In the special case where $K$ is diagonal, corresponding to a situation without entanglement in the ground state, $\Omega_{jj}= \sqrt{K_{jj}}$, and we recover the vacuum energy of the local oscillators, resulting in impossible work extraction.

The quantum heat is given by the mean energy difference before and after the measurement (see Table~\ref{table}),
\be
Q=\braket{H_\Loc}_{\tilde0}-E_{\tilde0}= \frac14\sum_j(K_{jj}(\Omega^{-1})_{jj}-\Omega_{jj}),
\label{qheat}
\ee
which is shown to be greater than or equal to 0 in Appendix~\ref{Qheatpos}. The efficiency is given by
\be
\eta=\frac{W}{W+\Delta},
\ee
where $\Delta=E_{0_\Loc}-E_{\tilde0}$ is again the local entanglement gap. A generalized expression for the generating function of all local energies can be found of Gaussian form in the generating variables. The probabilities are given explicitly in terms of a matrix Hafnian in Ref.~\cite{hamilton2017gaussian}.

The work output standard deviation~$\sigma$ can also be calculated exactly using similar techniques. Using the relation in Table~\ref{table}, we eventually find
\be
\sigma^2=\frac18\left(\sum_{j,k}K_{jj}(\Omega^{-1})_{jk}^2K_{kk}-\sum_jK_{jj}\right).
\label{sigma-Nosc}
\ee

In the case of two oscillators discussed in the previous section, we have
\be
K=
\begin{pmatrix}
k_0+g&-g\\
-g&k_0+g
\end{pmatrix},
\ee
Inserting these formulas into Eqs.~\eqref{mainresult} and~\eqref{sigma-Nosc} respectively recovers the expression for $\braket{H_\Loc}_{\tilde0}$ from Eq.~\eqref{Eloc2} and the expression for $\sigma$ from Eq.~\eqref{sigma-2osc}.

\subsection{Linear oscillator chain}

We consider a simple model of a linear chain of masses on springs with nearest-neighbor coupling to see how the work and efficiency scale with $N$. The symmetric, tridiagonal coupling matrix is given by $K=k_0(2 I -T)$, where $T$ has $1$s on the first off-diagonals and $0$s on the main diagonal. In this case, the local expected energy takes the form
\be
\braket{H_\Loc}_{\tilde0}=\frac{k_0}2\Tr(\Omega^{-1})+\frac14\Tr\Omega,
\label{mrsimp}
\ee
The tridiagonal coupling matrix $K$ can be diagonalized exactly~\cite{meyer2000matrix}, with eigenvalues
\be
k_m=2k_0\left(1-\cos\left(\frac{m\pi}{N+1}\right)\right)=4k_0\sin^2\left(\frac{m\pi}{2(N+1)}\right),
\ee
where $m=1,\dots,N$ delineates the mode number. The corresponding natural frequencies are
\be
\Omega_m=\sqrt{k_m}=2\sqrt{k_0}\sin\left(\frac{m\pi}{2(N+1)}\right).
\ee
Thus, we can express the local energy as
\be
\braket{H_\Loc}_{\tilde0}=\frac{k_0}2\sum_m\frac1{\Omega_m}+\frac14\sum_m\Omega_m.
\label{Hloc_Nosc}
\ee
The second sum in Eq.~\eqref{Hloc_Nosc} above can be calculated exactly:
\be
\sum_m\Omega_m=\sqrt{k_0}\left(\cot\left(\frac\pi{4(N+1)}\right)-1\right),
\ee
but the first one does not have a closed form solution. Therefore, we examine the limit~$N\to\infty$ and approximate the sum as an integral. We find the asymptotic behavior:
\be
\braket{H_\Loc}_{\tilde0}\underset{N\to\infty}\simeq\frac{N\sqrt{k_0}}{2\pi}\left(\ln\left(\frac{4 N}\pi\right)+C\right),
\label{eloc-asym}
\ee
where $C$ is a constant whose value can be estimated using the Euler-Maclaurin formula. The correction due to the first term in the formula (half the sum of the boundary values) yields $C=5/2$, when the exact numerical value is $C\approx2.577$.

The vacuum reference energies are given approximately as
\begin{align}
&E_{0_\Loc}\underset{N\to\infty}\simeq N\sqrt{\frac{k_0}2},\\
&E_{\tilde0}\underset{N\to\infty}\simeq\frac{2N\sqrt{k_0}}\pi.
\end{align}
Importantly, the local energy grows logarithmically as $N\ln N$ with $N$, while both vacuum energies scale only linearly with $N$, so the efficiency $\eta=W/Q$ goes to unity in the limit of large $N$. This effect comes from the fact that $\Omega_m\simeq m\pi\sqrt{k_0}/N$ for small $m/N$, so the sum of inverse frequencies has a logarithmic behavior in $N$. 

To assess the behavior of fluctuations, we examine the work output standard deviation. We have
\be
\sigma^2=\frac{k_0^2}2\Tr(K^{-1})-\frac{Nk_0}4.
\label{sigma-chain}
\ee
The trace in Eq.~\eqref{sigma-chain} above reduces to
\be
\Tr(K^{-1})=\sum_m\frac1{k_m}=\frac{N(N+2)}{6k_0}.
\ee
As a result, the standard deviation reads as
\be
\sigma=\frac{\sqrt{N(N-1)k_0}}{2\sqrt3}.
\ee
This indicates that fluctuations of the work output scale linearly with $N$ and are in consequence logarithmically suppressed when compared to the average work output as $N$ increases.

\begin{figure}
\centering
\includegraphics[width=\linewidth]{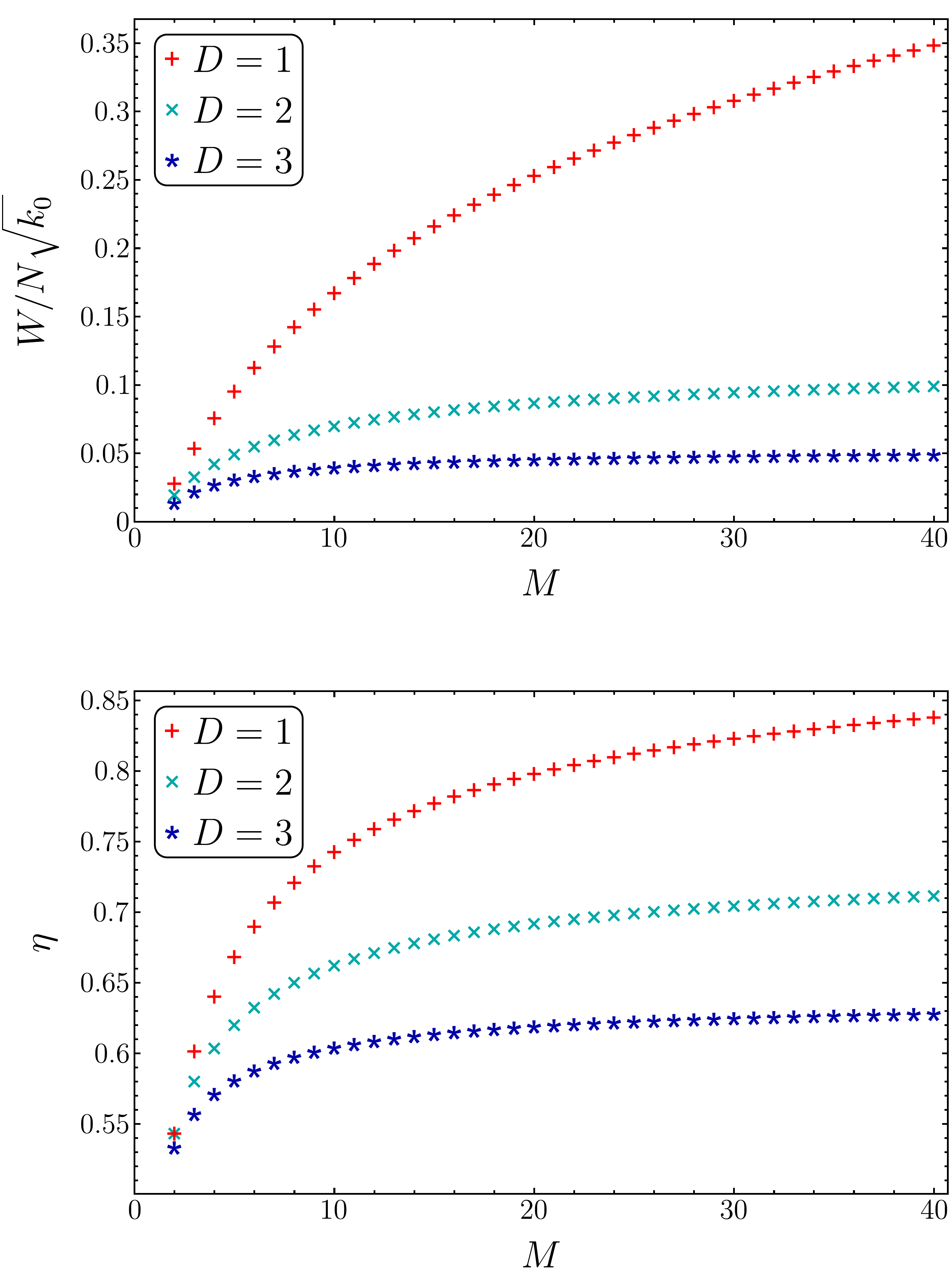}
\caption{Work per oscillator (upper panel) and efficiency (lower panel) for a $D$-dimensional cubical lattice with as a function of the number of oscillators~$M$ on a side. Red plusses, cyan crosses and blue asterisks correspond to $D=1$, $D=2$ and $D=3$ respectively.}
\label{weffvsd}
\end{figure}

The logarithmic enhancement is special to one dimension; in higher dimensions the low-frequency divergence is regularized and all energies typically scale linearly with $N$, so, similarly to the fermionic case, the efficiency is independent of $N$ in the large-$N$ limit. The work extracted from the oscillator network and the engine efficiency are shown in Fig.~\ref{weffvsd}. We show these quantities for a $D$-dimensional cubical geometry, where $M$ is the number of oscillators on the side, so $N=M^D$. For a rectangular geometry in $D$ dimensions with nearest-neighbor coupling~$k_0$, the frequency $\Omega_{\vec m}$ of mode number vector $\vec m=(m_1,\dots,m_D)$, where $m_j$ ranges from $1$ to $M$, is given by
\be
k_{\vec m}=\Omega_{{\vec m}}^2 = 4k_0\sum_{\alpha=1}^D\sin^2\left(\frac{m_\alpha\pi}{2(M+1)}\right).
\ee
We see in Fig.~\ref{weffvsd} that the work per oscillator saturates to a constant for $D=2$ and $D=3$, but continues to grow for $D=1$. As we move to higher dimensions, both the work per oscillator and the efficiency decrease. However, because the total work scales as $N=M^D$ in two and three dimensions, the work extraction grows exponentially with dimension when the number of oscillators on a side is kept fixed.

\section{Conclusion}
\label{ccl}

In this work, we forge connections between the field of quantum energetics, focusing on applications such as measurement-driven quantum engines, and that of quantum materials, which is concerned with topics such as quantum magnetism and phase transitions. We presented a protocol to extract work out of the quantum vacuum through local operations on a many-body system. To do so, we introduced an engine cycle during which the entangled ground state of an interacting many-body system is first destroyed as local measurements are carried out on each subsystem. This projects the system onto the local eigenbasis where interactions are turned off at no energetic cost. Work is then extracted by applying local feedback operations to each subsystem found in a local excited state. With the system now in its local ground state, interactions are turned back on at no energetic cost. Finally, the many-body system is put in contact with a cold bath so that it relaxes to its entangled ground state, and the cycle can restart. We note that an ``always on'' operation of the engine is possible, where the interaction couplings are not controlled during the engine cycle, however, the local controls must be much faster than any system dynamics and applied immediately after the measurement step, which will likely be experimentally challenging. 

We assessed the work output and efficiency of the cycle for various examples of interacting many-body systems. We first considered the simple case of two coupled qubits where work and efficiency can be calculated straightforwardly, and a clear trade-off between these quantities can be identified: maximum efficiency corresponds to no work output and vice versa. The working principle of this two-qubit engine can be extended to an arbitrarily long chain. Our model is analogous to the one-dimensional transverse-field Ising model which can be mapped to free fermions. This mapping yields exact results for the qubit chain engine's work and efficiency. The transverse-field Ising model is known to undergo a quantum phase transition which has a clear impact on the engine's performance: Both work and efficiency abruptly increase at the critical point. As a consequence, efficiency reaches its maximum value close to the critical point where the work output is nonzero.

We also analyzed the bosonic case with vacuum engines made from coupled oscillators. We considered two coupled oscillators and proceeded to analyze arbitrary oscillator networks. Analytical results can be obtained for work and efficiency in the general case. Cubical geometries with nearest-neighbor coupling in one, two and three dimensions were treated explicitly. High efficiencies can be achieved as the number of coupled oscillators increases. Remarkably, the efficiency approaches unity for a one-dimensional chain with nearest-neighbor couplings.

Throughout this work, fluctuations were calculated using the standard deviation for the work output. In both many-qubit and many-oscillator systems, we note that fluctuations tend to become negligible in comparison to averages in the limit of a large number of coupled subsystems. The engine's performance thus seems optimal in every way in this limit: The average work output increases linearly (or faster) with (at worst) a stagnation of efficiency while fluctuations become less and less significant.

In usual quantum computing platforms, local measurement is the comparatively easy part, while creating and maintaining entanglement is difficult in practice. Here the situation is reversed: entanglement in a many-body system comes ``for free'' by simply waiting for the system to relax to its ground state. The challenging part is to make projective measurements that are fast and in the local energy basis. Our analysis indicates this is possible, but the coupling strength of the meter to the local system must overwhelm the coupling to the neighboring quantum systems by an order of magnitude, at least. Since the amount of energy transferred to the system is, on average, equal to the quantum heat, some fraction of the energy used to turn on the coupling of the meter to the system can in principle be reused, so this resource should be viewed as a catalyst, rather than as part of the engine's fuel. The strong meter-coupling feature is the outstanding experimental challenge to implement this quantum vacuum engine in the laboratory.

\begin{acknowledgments}
This work was supported by the John Templeton Foundation, Grant No. 61835. A.A. acknowledges support from the National Research Foundation, Singapore and A*STAR under its CQT Bridging Grant, the Foundational Questions Institute Fund (Grants No. FQXi-IAF19-01 and No. FQXi-IAF19-05), and the ANR Research Collaborative Project “Qu-DICE” (Grant No. ANR-PRC-CES47).
\end{acknowledgments}

\bibliography{refs}

\appendix

\section{Cycle dynamics for the two-qubit engine}
\label{2qb_dynamics}

In this appendix, we analyze in more detail the dynamics of the two-qubit engine cycle. This will enable us to gauge the cycle's duration, and then estimate the engine's power output. We will be focusing on modeling the measurement and relaxation steps of the cycle.

\subsection{Measurement dynamics}

To investigate the dynamics of the local measurement procedure, we have to introduce a specific model for the measurement device. Since the measurement we are considering only has two possible outcomes, the measurement device can be modeled by an additional qubit, hereafter denoted by $\M$. Qubits~$\A$ and~$\M$ are coupled together such that the state of the former is imprinted on the latter. We then have to add a new coupling Hamiltonian into the picture:
\be
  H_\M=g_\M\sigma_\A^+\sigma_\A^-\sigma_\M^x,
  \label{HM}
\ee
where $g_\M$ corresponds to the measurement strength. The above Hamiltonian corresponds to the situation where the meter qubit~$\M$ is initially in its ground state~$\ket{0_\M}$. If qubit~$\A$ is in state~$\ket{1_\A}$, qubit~$\M$ flips, while nothing changes if qubit~$\A$ in state~$\ket{0_\A}$. The coupling Hamiltonian~$H_\M$ is turned on at time~$t=0$, and turned off at time~$t=t_\M$ when, ideally, the state of qubit~$\M$ is identical to that of qubit~$\A$ which can then be read out.

The joint state of qubits~$\A$ and~$\B$ at time~$t=0$ is the two-qubit ground state: $\ket{\phi^-}=\cos\phi\ket{00}-\sin\phi\ket{11}$. As such, the initial three-qubit state reads as
\be
  \ket{\Psi_0}=\cos\phi\ket{000}-\sin\phi\ket{110},
  \label{3qb_t0}
\ee
where the state of qubit~$\M$ is written after those of qubits~$\A$ and~$\B$. The Schr\"odinger equation governing the evolution of the three-qubit system under Hamiltonian~$H_\Loc+H_\Int+H_\M$ projected onto the computational basis yields
\be
\begin{cases}
\ii\dot\Psi_{000}=g\Psi_{110}/2,\\
\ii\dot\Psi_{001}=g\Psi_{111}/2,\\
\ii\dot\Psi_{010}=\omega_\mathrm B\Psi_{010}+g\Psi_{100}/2,\\
\ii\dot\Psi_{011}=\omega_\mathrm B\Psi_{011}+g\Psi_{101}/2,\\
\ii\dot\Psi_{100}=\omega_\mathrm A\Psi_{100}+g\Psi_{010}/2+g_\M\Psi_{101},\\
\ii\dot\Psi_{101}=\omega_\mathrm A\Psi_{101}+g\Psi_{011}/2+g_\M\Psi_{100},\\
\ii\dot\Psi_{110}=(\omega_\mathrm A+\omega_\mathrm B)\Psi_{110}+g\Psi_{000}/2+g_\M\Psi_{111},\\
\ii\dot\Psi_{111}=(\omega_\mathrm A+\omega_\mathrm B)\Psi_{111}+g\Psi_{001}/2+g_\M\Psi_{110}.
\end{cases}
\label{SchrEq_3qb}
\ee
One can observe that the four outer equations in the above system are decoupled from the four inner ones. Furthermore, all the components appearing in these inner equations are zero in the initial state given in Eq.~\eqref{3qb_t0}. We deduce that they will stay as such at all times: $\Psi_{010}(t)=\Psi_{011}(t)=\Psi_{100}(t)=\Psi_{101}(t)=0$. It is cumbersome but straightforward to solve the remaining coupled differential equations. One eventually finds that the nonzero components~$\Psi_{000}$,~$\Psi_{001}$,~$\Psi_{110}$, and~$\Psi_{111}$ oscillate with characteristic frequencies~$\Omega_{pq}$, $p,q=\pm$, given by
\be
\Omega_{pq}=\frac{\omega_\A+\omega_\B}2\left(1+p\gamma_\M+q\sqrt{(1+p\gamma_\M)^2+\gamma^2}\right),
\ee
where $\gamma=g/(\omega_\A+\omega_\B)$ and $\gamma_\M=g_\M/(\omega_\A+\omega_\B)$. More precisely, the solution to the system in Eq.~\eqref{SchrEq_3qb} reads as
\begin{align}
&\Psi_{000}(t)=\frac12\sum_{p,q=\pm}\frac1{1+r_{pq}^2}(\cos\phi-r_{pq}\sin\phi)\e^{-\ii\Omega_{pq}t},\\
&\Psi_{001}(t)=\frac12\sum_{p,q=\pm}\frac p{1+r_{pq}^2}(\cos\phi-r_{pq}\sin\phi)\e^{-\ii\Omega_{pq}t},\\
&\Psi_{110}(t)=\frac12\sum_{p,q=\pm}\frac{r_{pq}}{1+r_{pq}^2}(\cos\phi-r_{pq}\sin\phi)\e^{-\ii\Omega_{pq}t},\\
&\Psi_{111}(t)=\frac12\sum_{p,q=\pm}\frac{pr_{pq}}{1+r_{pq}^2}(\cos\phi-r_{pq}\sin\phi)\e^{-\ii\Omega_{pq}t},
\end{align}
with $r_{pq}=2\Omega_{pq}/g$.

\begin{figure}
\centering
\includegraphics[width=\linewidth]{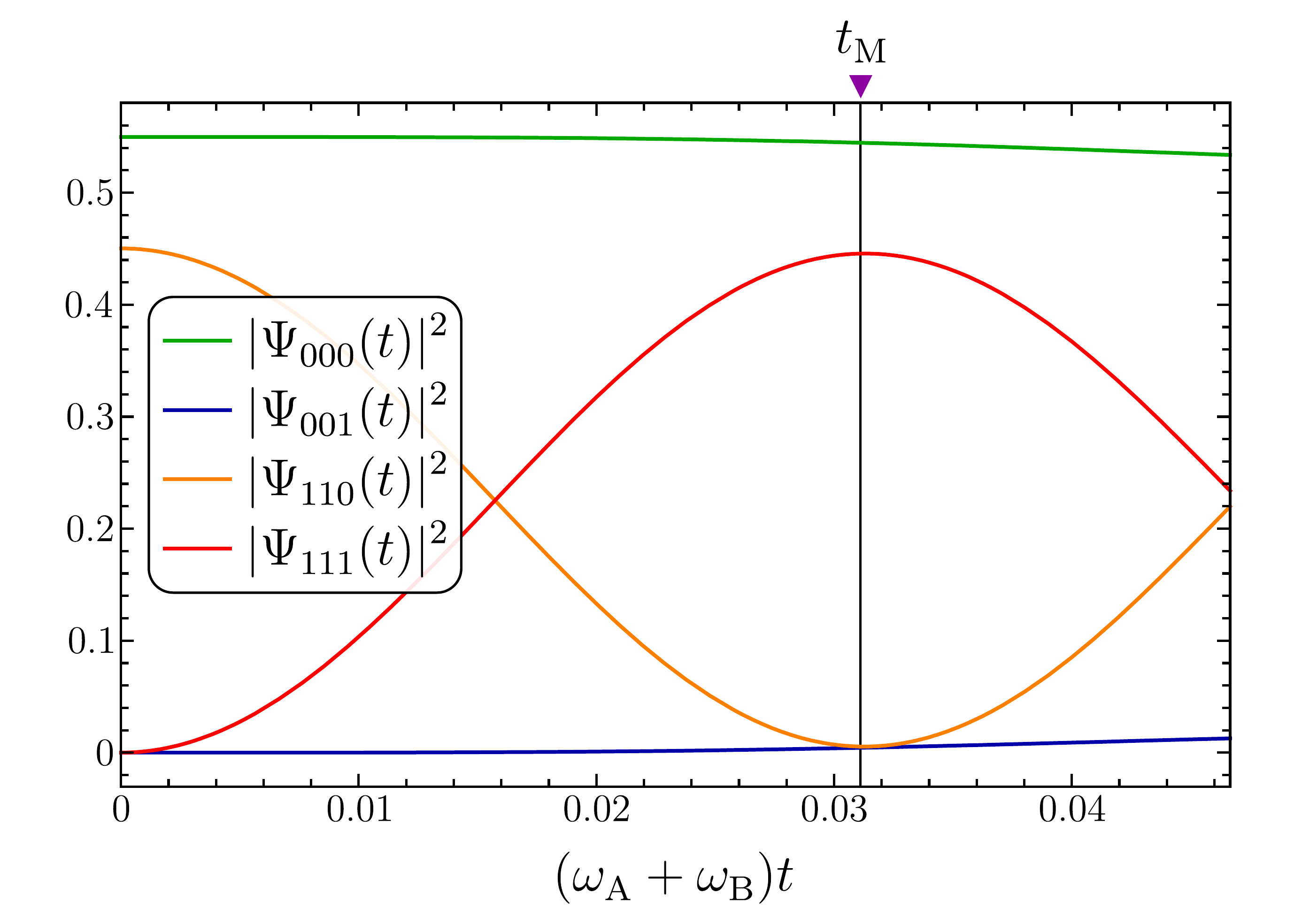}
\caption{Plot of the probabilities~$\abs{\Psi_{000}(t)}^2$, $\abs{\Psi_{001}(t)}^2$, $\abs{\Psi_{110}(t)}^2$ and~$\abs{\Psi_{111}(t)}^2$. The interaction Hamiltonian~$H_\M$ in Eq.~\eqref{HM} is turned on at $t=0$. Measurement should be carried out at time~$t_\M$ after half an oscillation of the probabilities~$\abs{\Psi_{110}}^2$ (orange curve) and~$\abs{\Psi_{111}}^2$ (red curve). Conversely, the probabilities~$\abs{\Psi_{000}}^2$ (green curve) and~$\abs{\Psi_{001}}^2$ (blue curve) are almost constant between times~$0$ and~$t_\M$. The parameters for this plot are $g=10(\omega_\A+\omega_\B)$ and $g_\M=50(\omega_\A+\omega_\B)$.}
\label{MeasProba}
\end{figure}

Ideally, the coupling Hamiltonian~$H_\M$ should be turned off at a time~$t_\M$ when the readout of qubit~$\M$ matches the initial joint state of qubits~$\A$ and~$\B$, that is, $\abs{\Psi_{000}(t_\M)}^2=\cos^2\phi$ and $\abs{\Psi_{111}(t_\M)}^2=\sin^2\phi$, while $\Psi_{001}(t_\M)=\Psi_{110}(t_\M)=0$. However, the existence of such a perfectly accurate measurement is not guaranteed in realistic models such as the one considered here. Instead, we estimate the time necessary to perform the measurement as accurately as possible by analyzing the population exchange between the three-body states~$\ket{110}$ and~$\ket{111}$: The coupling between qubits~$\A$ and~$\M$ flips the latter if the former is in its excited state thus inverting the probabilities~$\abs{\Psi_{110}(t)}^2$ (nonzero at $t=0$) and~$\abs{\Psi_{111}(t)}^2$ (zero at $t=0$). Such an inversion peaks, roughly speaking, after half a period of the term that oscillates the fastest in $\abs{\Psi_{110}(t)}^2$ and $\abs{\Psi_{111}(t)}^2$. Since these quantities can be expressed as sums of cosines with frequencies~$\Omega_{pq}-\Omega_{p'q'}$, we estimate~$t_\M$ as $t_\M=\pi/\nu$, where $\nu=\max_{p,p',q,q'}\abs{\Omega_{pq}-\Omega_{p'q'}}$ is the largest Bohr frequency. We find
\be
\nu=\Omega_{++}-\Omega_{--}=\frac{\omega_\A+\omega_\B}2\Big(
\begin{aligned}[t]
&2\gamma_\M+\sqrt{(1+\gamma_\M)^2+\gamma^2}\\&+\sqrt{(1-\gamma_\M)^2+\gamma^2}\Big).
\end{aligned}
\ee
This estimate yields accurate results when the difference between $\nu$ and $\Omega_{++}-\Omega_{+-}$ (the second largest Bohr frequency) is large enough so that oscillations at slower frequencies can be neglected. One should also note that $\abs{\Psi_{000}(t)}^2$ and $\abs{\Psi_{001}(t)}^2$ remain roughly constant throughout the measurement process which is why their dynamics have not been considered in the analysis above. These observations are substantiated in Fig.~\ref{MeasProba}.

\begin{figure}
  \centering
  \includegraphics[width=\linewidth]{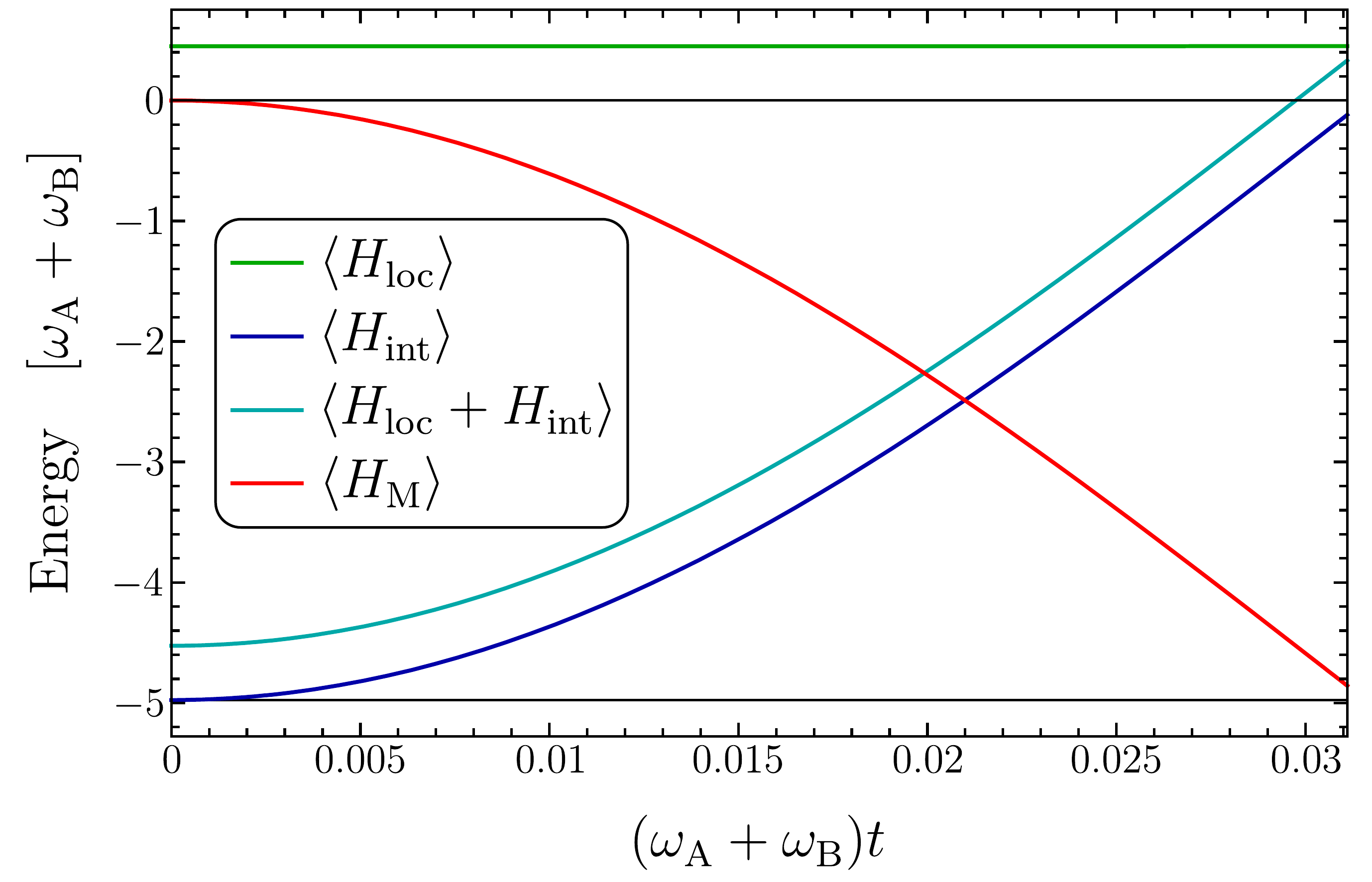}
  \caption{Energy transfers between the two-qubit system and the measurement qubit~$\M$. The two-qubit average energy~$\braket{H_\Loc+H_\Int}$ (cyan curve) clearly increases at the expense of $\braket{H_\M}$ (red curve) which shows that energy is transferred from qubit~$\M$ to the two-qubit system. More precisely, the entanglement between qubits~$\A$ and~$\B$ is destroyed by the local measurement. The corresponding binding energy~$\braket{H_\Int}$ (blue curve) consequently vanishes while the local contribution to the two-qubit energy~$\braket{H_\Loc}$ (green curve) is barely affected by the measurement. The parameters for this plot are $g=10(\omega_\A+\omega_\B)$ and $g_\M=50(\omega_\A+\omega_\B)$.}
  \label{MeasEnergy}
\end{figure}

Furthermore, our approach also enables us to assess more rigorously the energy transfers between qubit~$\M$ and qubits~$\A$ and~$\B$. As shown in Fig.~\ref{MeasEnergy}, the energy for the two-qubit system~$\braket{H_\Loc+H_\Int}$ increases at the expense of the measurement energy~$\braket{H_\M}$. This is almost solely caused by the interaction term~$\braket{H_\Int}$ vanishing, which mirrors the decrease of $\braket{H_\M}$. The local measurement operation indeed destroys the entanglement between qubits~$\A$ and~$\B$. The corresponding binding energy thus supplied to the joint system is then used as a resource for work extraction.

\subsection{Relaxation dynamics}

We now tackle the relaxation step of the two-qubit engine cycle. We model it by assuming that qubits~$\A$ and~$\B$ are coupled to the same bath with which they can individually exchange photons. The Hamiltonian for the bath is simply given by
\be
H_\mathrm{bath}=\sum_\lambda\ce_\lambda a_\lambda^\dagger a_\lambda,
\ee
where $a_\lambda^\dagger$ and $a_\lambda$ respectively create and annihilate a photon in mode~$\lambda$ in the bath, with $\ce_\lambda$ the corresponding energy. The system-bath interaction Hamiltonian reads as
\be
V=\sum_{j,\lambda}\kappa_\lambda\sigma_j^+a_\lambda+\text{h.c.},
\ee
which corresponds to the situation where both qubits are coupled with mode~$\lambda$ in the bath with the same amplitude~$\kappa_\lambda$. We describe the dynamics of the two-qubit system coupled to the bath using the Gorini--Kossakowski--Sudarshan--Lindblad (GKSL) formalism. Such an approach is valid provided that the bath correlation time~$\tau_\mathrm{corr}$ is much shorter than the damping time~$\tau_\mathrm{damp}$, defined as the typical time scale over which the system state changes noticeably. This condition is generally satisfied when the bath is weakly coupled to the system and has a broad density of states, which ensures that excitations in the bath resulting from its interaction with the system decay quickly. The GKSL formalism further relies on a secular approximation which holds when the typical time scale for the intrinsic evolution of the two-qubit system is much shorter than $\tau_\mathrm{damp}$. If the conditions above are met, the dynamics of the reduced density matrix for the two-qubit system~$\rho$ obeys the GKSL master equation:
\be
\diff\rho t=-\ii[H_\Loc+H_\Int+\Lambda,\rho]+\mathcal D[\rho],
\label{MEq}
\ee
where 
$\Lambda$ is the Lamb shift Hamiltonian and $\mathcal D[\rho]$ is the dissipator. In what follows we will focus on the dynamics resulting from the dissipator which describes the system's decay towards its thermal state, while the commutator in Eq.~\eqref{MEq} above will only give rise to additional phase factors in the density matrix off-diagonal elements.

\subsubsection{Dynamics of populations}

It is well-known that the master equation takes the form of a rate equation for the diagonal elements of the density matrix (in the system's eigenbasis):
\be
\diff{p_\chi}t=\sum_{\chi'\ne\chi}(\Gamma_{\chi\chi'}p_{\chi'}-\Gamma_{\chi'\chi}p_\chi),
\ee
where $p_\chi=\braket{\chi|\rho|\chi}$, and $\Gamma_{\chi\chi'}$ denotes the rate for the transition from eigenstate~$\ket{\chi'}$ to eigenstate~$\ket\chi$. These transition rates satisfy the detailed balance condition:
\be
  \Gamma_{\chi'\chi}=\Gamma_{\chi\chi'}\e^{\beta(\omega_\chi-\omega_{\chi'})},
  \label{DetailedBalance}
\ee
where $\beta=(k_\mathrm BT)^{-1}$ is the inverse temperature for the bath. The detailed balance condition implies that the steady-state populations follow the Boltzmann distribution:
\be
  p_\chi(t\to\infty)=\frac1Z\e^{-\beta\omega_\chi}.
\ee
Here, we are interested in having the two-qubit system relaxing to its ground state when put in contact with the bath. Its temperature should then be low enough so that the first excited state's population is negligible when compared to that of the ground state. Considering the eigenenergies in Eqs.~\eqref{wphi} and~\eqref{wpsi}, this condition becomes
\be
  k_\mathrm BT\ll\sqrt{1+\gamma^2}-\sqrt{\delta^2+\gamma^2}.
  \label{lowT}
\ee
Further, half of the transition rates~$\Gamma_{\chi\chi'}$ can be neglected in this low-temperature limit as a consequence of the detailed balance condition in Eq.~\eqref{DetailedBalance}. More precisely, all rates corresponding to an increase in the two-qubit system's energy are neglected since these transitions involve the bath emitting a photon to excite one of the qubits, which becomes exponentially unlikely at low temperature. The master equation can then be written as follows
\be
  \diff*{
  \begin{pmatrix}
      p_\phi^+\\p_\psi^+\\p_\psi^-\\p_\phi^-
  \end{pmatrix}
  }t=-\begin{pmatrix}
      \Gamma_++\Gamma_-&0&0&0\\
      -\Gamma_+&\Gamma_+&0&0\\
      -\Gamma_-&0&\Gamma_-&0\\
      0&-\Gamma_+&-\Gamma_-&0
  \end{pmatrix}
  \begin{pmatrix}
      p_\phi^+\\p_\psi^+\\p_\psi^-\\p_\phi^-
  \end{pmatrix}
  \label{MEq_lowT}
\ee
with
\begin{align}
  &\Gamma_+=\pi K\left(1+\frac1{\sqrt{1+\gamma^2}}\right)\left(1+\frac\gamma{\sqrt{\delta^2+\gamma^2}}\right),\\
  &\Gamma_-=\pi K\left(1+\frac1{\sqrt{1+\gamma^2}}\right)\left(1-\frac\gamma{\sqrt{\delta^2+\gamma^2}}\right),
\end{align}
where $K$ denotes the spectral density for the bath. It is defined as $K=\sum_\lambda\abs{\kappa_\lambda}^2\delta(E-\ce_\lambda)$, and it is assumed to be energy-independent for simplicity.

We are not interested here in the explicit solution to Eq.~\eqref{MEq_lowT}, but solely in the typical relaxation time for the two-qubit system's relaxation. It can be estimated by considering the eigenvalues of the transition rate matrix in Eq.~\eqref{MEq_lowT}; more precisely, we are interested in the smallest non-zero eigenvalue which is the inverse of the longest characteristic time for the decay of populations. The calculation is straightforward, and we find that the aforementioned characteristic time is
\be
  t_\mathrm p=\frac1{\Gamma_-}=\frac1{\pi K}\left(1+\frac1{\sqrt{1+\gamma^2}}\right)^{-1}\left(1-\frac\gamma{\sqrt{\delta^2+\gamma^2}}\right)^{-1}.
  \label{tp}
\ee

\subsubsection{Dynamics of coherences}

In the situation at stake here, the two-qubit density matrix at the beginning of the relaxation phase is $\rho_0=\ket{00}\bra{00}$. This means that there are initial coherences in the $(\ket{\phi^+},\ket{\psi^+},\ket{\psi^-},\ket{\phi^-})$~eigenbasis, namely, $\braket{\phi^+|\rho_0|\phi^-}=\braket{\phi^-|\rho_0|\phi^+}=\sin\phi\cos\phi$ while all other off-diagonal elements are zero. The master equation does not couple $\braket{\phi^+|\rho|\phi^-}$ and $\braket{\phi^-|\rho|\phi^+}$ with other off-diagonal elements as a consequence of the secular approximation. We then find that the decay of coherences due to the coupling to the bath boils down to
\be
\abs{\braket{\phi^+|\rho(t)|\phi^-}}=\abs{\braket{\phi^-|\rho(t)|\phi^+}}=\frac{\gamma\e^{-t/t_\mathrm c}}{2\sqrt{1+\gamma^2}},
\ee
with 
\be
  t_\mathrm c=\frac2{\Gamma_++\Gamma_-}=\frac1{\pi K}\left(1+\frac1{\sqrt{1+\gamma^2}}\right)^{-1}.
\ee

As is usually the case, we find that coherences decay faster than populations, that is, $t_\mathrm p>t_\mathrm c$; $t_\mathrm p$ then defines the relevant time scale to estimate the two-qubit system's relaxation time.

\subsection{Power output of the engine}

In the previous subsections, we have evaluated the durations of the measurement and relaxation steps of our cycle. Assuming that the remaining part of the cycle (applying local pulses to each qubit) be carried out almost instantaneously, the power output of the two-qubit engine can be estimated as
\be
  P\approx\frac W{t_\M+5t_\mathrm p},
\ee
where the factor of $5$ in the denominator corresponds to a probability of $99\%$ to find the two-qubit system in its ground state after the relaxation step. Remarkably, while the work output of the engine only depends on the sum~$\omega_\A+\omega_\B$, see Eq.~\eqref{work}, its power output also depends on the detuning~$\delta$ through the characteristic relaxation time~$t_\mathrm p$, see Eq.~\eqref{tp}. As $t_\mathrm p$ decreases with $\delta$, the same amount of work can be extracted faster for larger detunings. However, larger detunings also correspond to situations where the two lowest eigenenergies~$E_\phi^-$ and~$E_\psi^-$ are closer to one another, see Eqs.~\eqref{wphi} and~\eqref{wpsi}. It then becomes increasingly challenging to ensure that the temperature of the bath is low enough so that the two-qubit system indeed relaxes to its ground state as shown in Eq.~\eqref{lowT}.

\section{Asymptotic results for an open chain}
\label{OpenChain}

In this appendix, we analyze the case of a chain of $N$ coupled qubits with open boundary conditions. We then consider the Hamiltonian
\be
H=H_\Loc+H_\Int=\sum_{j=1}^N\omega_j\sigma_j^+\sigma_j^-+\frac12\sum_{j=1}^{N-1}g_j\sigma_j^x\sigma_{j+1}^x,
\ee
which describes $N$ qubits whose transitions frequencies are denoted by $\omega_j$, $j=1,\dots,N$, coupled to their nearest neighbors, where $g_j$ corresponds to the coupling amplitude between sites~$j$ and~$j+1$.

While this model can be solved exactly using the Jordan--Wigner transformation, we will focus here on the weak and deep strong coupling limits for simplicity.

\subsection{Weak coupling limit}

We first analyze the weak coupling limit where $H_\Int$ can be treated as a perturbation with respect to $H_\Loc$. The eigenstates of the local Hamiltonian are separable and can thus be written as $\ket l=\ket{l_1,\dots,l_N}$, where $l_j=0,1$ represents the state of qubit~$j$. A given local eigenstate~$\ket l$ corresponds to the representation of the binary number
\be
  l=\sum_{j=1}^N2^{N-j}l_j,
\ee
where $0\le l\le 2^N-1$. We then have:
\be
  H_\Loc\ket l=\sum_{j=1}^Nl_j\omega_j\ket l.
\ee
The ground state is obviously $\ket{0_\Loc}=\ket{l=0}=\ket{0,\dots,0}$ with the corresponding eigenenergy set at $0$. We now apply perturbation theory to obtain the leading-order corrections to the ground state and its energy.

To first order in perturbation, there is no correction to the energy; however, the ground state becomes
\be
  \ket{\tilde0}\simeq\ket{0_\Loc}-\frac12\sum_{j=1}^{N-1}\frac{g_j}{\omega_j+\omega_{j+1}}\ket{1_j1_{j+1}},
\ee
where the state~$\ket{1_j1_{j+1}}$ denotes the many-body state where only the qubits at positions~$j$ and~$j+1$ are in their excited state; it can also be written as $\ket{l=3\times2^{N-j-1}}$. To second order in perturbation, we find that the local entanglement gap is given by
\be
\Delta\simeq\frac14\sum_{j=1}^{N-1}\frac{g_j^2}{\omega_j+\omega_{j+1}},
\ee
and an additional correction to the ground state arises:
\be
  \braket{0_\Loc|\tilde0}\simeq1-\frac18\sum_{j=1}^{N-1}\left(\frac{g_j}{\omega_j+\omega_{j+1}}\right)^2.
\ee

Performing local measurements on each qubit within the chain while it is in its ground state can yield the following outcomes: one either obtains the local ground state~$\ket{0_\Loc}$, in which case no energy can be extracted, or one finds that the qubits at positions~$j$ and~$j+1$ are in their excited state. In the latter case, an amount of energy~$\omega_j+\omega_{j+1}$ can be extracted by applying local pulses on these two qubits. The corresponding probabilities for these outcomes are
\begin{align}
  &P_{0_\Loc}\simeq1-\frac14\sum_{j=1}^{N-1}\left(\frac{g_j}{\omega_j+\omega_{j+1}}\right)^2,\\
  &P_{1_j1_{j+1}}\simeq\frac14\left(\frac{g_j}{\omega_j+\omega_{j+1}}\right)^2.
\end{align}
We then deduce the average work output for the engine:
\be
  W=\braket{\tilde0|H_\Loc|\tilde0}\simeq\frac14\sum_{j=1}^{N-1}\frac{g_j^2}{\omega_j+\omega_{j+1}}=\Delta,
\ee
and the efficiency is given by
\be
  \eta=\frac{W}{W+\Delta}=\frac12.
\ee

One should note that, when all transition frequencies and nearest-neighbor coupling parameters are equal, $\omega_j=\omega$ and $g_j=g$ for all $j$, the work output scales linearly with $N$ as it becomes
\be
  W=\frac{(N-1)g^2}{8\omega}.
\ee
Conversely, the efficiency is independent of $N$.

\subsection{Deep strong coupling limit}

We now consider the deep strong coupling limit where $H_\Loc$ is treated as a perturbation with respect to $H_\Int$. Denoting by $\ket\pm$ the eigenstates of $\sigma^x$,
\be
  \ket\pm=\frac1{\sqrt2}(\ket1\pm\ket0),
\ee
the eigenstates of $H_\Int$ can be written as $\ket\alpha$, where $\alpha=(\alpha_1,\dots,\alpha_N)$ is a multi-index such that $\alpha_j=\pm$. We then have
\be
  H_\Int\ket\alpha=\frac12\sum_{j=1}^N\alpha_j\alpha_{j+1}g_j\ket\alpha.
  \label{diag_Hint}
\ee
The ground state is clearly two-fold degenerate with the eigenstates~$\ket{\pm,\mp,\dots}$ and~$\ket{\mp,\pm,\dots}$. However, these states do not satisfy one of the symmetries of the total Hamiltonian~$H=H_\Loc+H_\Int$ and are therefore not the relevant ground states for our study. Indeed, let us consider the parity operator along the $x$ axis: $\sigma^z=\ket{+}\!\bra{-}+\ket{-}\!\bra{+}$. It is straightforward to check that $\sigma^z\sigma^+\sigma^-\sigma^z=\sigma^+\sigma^-$ and $\sigma^z\sigma^x\sigma^z=-\sigma^x$. Considering the global parity operator~$\Pi=\sigma_1^z\otimes\dots\otimes\sigma_N^z$, it is then straightforward to show that $\Pi H\Pi=H$. This indicates that the appropriate eigenstates of $H_\Int$ to consider should also be eigenstates of the parity operator~$\Pi$. We consequently define the ground states
\be
  \ket{\chi^\pm}=\frac1{\sqrt2}(\ket{+,-,\dots}\pm\ket{-,+,\dots}),
\ee
where
\begin{align}
  &H_\Int\ket{\chi^\pm}=-\frac12\sum_{j=1}^{N-1}g_j\ket{\chi^\pm},\\
  &\Pi\ket{\chi^\pm}=\pm\ket{\chi^\pm}.
\end{align}

One can apply perturbation theory to understand how the degeneracy between these states is lifted. However, this splitting will only happen to $N$th order in perturbation. Indeed, degeneracy is lifted when one can connect the two eigenstates by repeated applications of the perturbation Hamiltonian, $H_\Loc$ here. To connect $\ket{\chi^+}$ and $\ket{\chi^-}$, or $\ket{+,-,\dots}$ and $\ket{-,+,\dots}$, $H_\Loc$ must be applied $N$ times. This is because one application $H_\Loc$ flips one qubit along the $x$ direction while connecting $\ket{+,-,\dots}$ and $\ket{-,+,\dots}$ clearly requires flipping all the qubits within the chain. It is then increasingly challenging to obtain analytical results as the number of qubits increases. The state of positive parity~$\ket{\chi^+}$ is the actual ground state in the deep strong coupling limit, and we will stick to zeroth order in perturbation hereafter.

The engine's work output is given by the expectation value for the local Hamiltonian in the interacting ground state. To calculate this expectation value, we rewrite the interacting ground in the local eigenbasis. We obtain
\be
  \ket{\tilde0}\simeq\ket{\chi^+}=\frac1{2^{(N+1)/2}}\sum_{l=0}^{2^N-1}(-1)^{e(l)}\left(1+(-1)^{f(l)}\right)\ket l,
\ee
where $f(l)$ is the total number of $1$s in the binary representation for $l$, while $e(l)$ is the number of $1$s appearing at even positions in this representation. We then find that any state~$\ket l$ with an even number of qubits in their excited state is an equiprobable outcome for our local measurement, while it is impossible to obtain a state~$\ket l$ with an odd number of qubits in their excited state:
\be
  P_l=\frac{1+(-1)^{f(l)}}{2^N}=
  \begin{cases}
  2^{-N+1}&\text{if $f(l)$ is even,}\\
  0&\text{if $f(l)$ is odd.}
  \end{cases}
\ee
The amount of energy that can then be extracted by applying a local pulse to each excited qubit is $\sum_{j=1}^Nl_j\omega_j$. The average work output of the engine consequently reads as
\be
  W=\braket{\tilde0|H_\Loc|\tilde0}\simeq\frac12\sum_{j=1}^N\omega_j.
\ee
We deduce the entanglement gap from Eq.~\eqref{diag_Hint},
\be
\Delta\simeq\frac12\sum_{j=1}^{N-1}g_j,
\ee
such that the efficiency is given by
\be
  \eta=\frac{W}{W+\Delta}=\left(1+\frac{\sum_{j=1}^{N-1}g_j}{\sum_{j=1}^N\omega_j}\right)^{-1}\simeq\frac{\sum_{j=1}^N\omega_j}{\sum_{j=1}^{N-1}g_j}.
\ee

Again, we find that the work output scales linearly with $N$ when all transition frequencies and nearest-neighbor coupling parameters are equal, $\omega_j=\omega$ and $g_j=g$ for all $j$, contrary to the efficiency which is almost constant:
\begin{align}
  &W=\frac{N\omega}2,\\
  &\eta=\left(1+\frac{(N-1)g}{N\omega}\right)^{-1}\simeq\frac{N\omega}{(N-1)g}.
\end{align}

\section{Detailed solution for a closed chain}
\label{Details}

Using the Jordan--Wigner transformation in Eq.~\eqref{JordanWigner}, spin operators are transformed into fermionic ones:
\be
c_j=\exp\left(\ii\pi\sum_{k=1}^{j-1}\sigma_k^+\sigma_k^-\right)\sigma_j^-. 
\ee
The total Hamiltonian~$H=H_\Loc+H_\Int$ then becomes
\be
  H=\omega\sum_{j=1}^Nc_j^\dagger c_j+\frac g2\Bigg(
  \begin{aligned}[t]
  &\sum_{k=1}^{N-1}\left(c_j^\dagger-c_j\right)\left(c_{j+1}^\dagger+c_{j+1}\right)\\
  &-\e^{n\ii\pi}\left(c_N^\dagger-c_N\right)\left(c_1^\dagger+c_1\right)\Bigg),
  \end{aligned}
\ee
where $n=\sum_j\sigma_j^+\sigma_j^-=\sum_jc_j^\dagger c_j$ is the total number of excitations across the chain. We wish to write the Hamiltonian in the following translation-invariant form:
\be
  H=\sum_{j=1}^N\left(\omega c_j^\dagger c_j+\frac g2\left(c_j^\dagger-c_j\right)\left(c_{j+1}^\dagger+c_{j+1}\right)\right).
\ee
We then understand that different boundary conditions must be applied depending on the number of excitations: $c_{N+1}=-c_1$ (antiperiodic boundary conditions) if $n$ is even and $c_{N+1}=c_1$ (periodic boundary conditions) if $n$ is even, or, equivalently, $c_{N+1}=-\e^{n\ii\pi}c_1$.

We now move to momentum space and introduce
\be
  c_p=\frac1{\sqrt N}\sum_{j=1}^Nc_j\e^{-j\ii p}.
\ee
The set of relevant values for the momentum~$p$ depends on the aforementioned boundary conditions: We have $p=(2m-1)\pi/N$ for antiperiodic boundary conditions, and $p=2m\pi/N$ for periodic boundary conditions, where, in both cases, $m$ is an integer ranging from $-\lfloor(N-1)/2\rfloor$ to $\lfloor N/2\rfloor$. In either case, we obtain
\be
  H=\sum_p
  \begin{pmatrix}
      c_p^\dagger&c_{-p}
  \end{pmatrix}
  H_p
  \begin{pmatrix}
      c_p\\c_{-p}^\dagger
  \end{pmatrix}
  +\frac{N\omega}2,
\ee
where $H_p$ is defined as
\be
  H_p=
  \begin{pmatrix}
      \omega+g\cos p&\ii g\sin p\\
      -\ii g\sin p&-(\omega+g\cos p)
  \end{pmatrix}.
  \label{Hp}
\ee
This matrix is diagonalized as follows:
\be
\begin{pmatrix}
    u_p&\ii v_p\\
    \ii v_p&u_p
\end{pmatrix}
H_p
\begin{pmatrix}
    u_p&-\ii v_p\\
    -\ii v_p&u_p
\end{pmatrix}
=
\begin{pmatrix}
    \Omega_p&0\\
    0&-\Omega_p
\end{pmatrix}
\ee
with
\begin{align}
  &u_p=\frac{g\abs{\sin p}}{\sqrt{2\Omega_p(\Omega_p-\omega-g\cos p)}},\\
  &v_p=\frac{\sgn p}{\sqrt2}\sqrt{1-\frac{\omega+g\cos p}{\Omega_p}},\\
  &\Omega_p=\sqrt{\omega^2+g^2+2\omega g\cos p}.
\end{align}
We consequently perform the Bogoliubov transformation: $\zeta_p=u_pc_p+\ii v_pc_{-p}^\dagger$, and obtain
\be
  H=\sum_p\Omega_p\left(\zeta_p^\dagger\zeta_p-\frac12\right)+\frac{N\omega}2.
\ee
One should note that the matrix~$H_p$ in Eq.~\eqref{Hp} is already diagonal if $p$ is a multiple of $\pi$. In such a case, we then have $u_p=1$, $v_p=0$, and $\Omega_p=\omega+g\cos p$.

One can check that
\be
  \zeta_p(u_p-\ii v_pc_p^\dagger c_{-p}^\dagger)\ket{0_\Loc}=\zeta_{-p}(u_p-\ii v_pc_p^\dagger c_{-p}^\dagger)\ket{0_\Loc}=0,
\ee
where $\ket{0_\Loc}=\otimes_j\ket0$ denotes the local ground state where each qubit is in its individual ground state; it corresponds to the vacuum state for Jordan--Wigner fermions. Using this property, we can construct the quasiparticle vacuum~$\ket{\tilde0}$ as follows:
\be
  \ket{\tilde0}=\prod_p\left(\sqrt{u_p}-\frac{\ii v_p}{\sqrt{u_p}}c_p^\dagger c_{-p}^\dagger\right)\ket{0_\Loc}.
\ee
The quasiparticle vacuum is the many-body ground state for $H$. We clearly see that it is a linear combination of states with an even number of excitations. This determines the set of momentum values to consider for subsequent calculations: $p=(2m-1)\pi/N$, where $m$ is an integer between $-\lfloor(N-1)/2\rfloor$ and $\lfloor N/2\rfloor$.

\section{Proof that the quantum heat for the oscillator network engine is positive}
\label{Qheatpos}

In this appendix, we demonstrate that the quantum heat for the oscillator network engine in Eq.~\eqref{qheat} is positive. The quantum heat is given by
\be
Q=\frac14\sum_j\left(K_{jj}(\Omega^{-1})_{jj}+\Omega_{jj}\right).
\ee
with $K=O^\mathrm TK_\mathrm DO$, where $O$ is an orthogonal matrix and $K_\mathrm D$ is a diagonal matrix with positive eigenvalues, and $\Omega=O^\mathrm TK_\mathrm D^{1/2}O$ is the square root of $K$. We then write $(K_\mathrm D)_{jk}=\delta_{jk}\Omega_j^2$, which yields
\be
\Omega_{jj}=\sum_{k,l}O_{kj}(K_\mathrm D^{1/2})_{kl}O_{lj}=\sum_kO_{kj}^2\Omega_k.
\ee
Similarly, we have
\begin{align}
&K_{jj}=\sum_kO_{kj}^2\Omega_k^2,\\
&(\Omega^{-1})_{jj}=\sum_k\frac{O_{kj}^2}{\Omega_k}.
\end{align}
The quantum heat consequently reads as
\be
Q=\frac14\sum_j\left(\sum_{k,l}O_{kj}^2O_{lj}^2\frac{\Omega_k^2}{\Omega_l}-\sum_kO_{kj}^2\Omega_k\right).
\label{qheat-2}
\ee
Since $O$ is an orthogonal matrix, we have
\be
(O^\mathrm TO)_{jj}=1=\sum_kO_{kj}^2,
\ee
which we insert into the last term to the right-hand side of Eq.~\eqref{qheat-2} to obtain
\be
Q=\frac14\sum_{j,k,l}O_{kj}^2O_{lj}^2\left(\frac{\Omega_k^2}{\Omega_l}-\Omega_k\right).
\ee
Swapping the dummy indices~$k$ and~$l$, this can be rewritten in the symmetric form
\be
Q=\frac18\sum_{j,k,l}O_{kj}^2O_{lj}^2\left(\frac{\Omega_k^2}{\Omega_l}+\frac{\Omega_l^2}{\Omega_k}-(\Omega_k+\Omega_l)\right).
\ee
Finally, we have
\be
\frac{\Omega_k^2}{\Omega_l}+\frac{\Omega_l^2}{\Omega_k}-(\Omega_k+\Omega_l)=\frac{(\Omega_k-\Omega_l)^2(\Omega_k+\Omega_l)}{\Omega_k\Omega_l}\ge0,
\ee
which concludes the proof that $Q\ge0$.

\end{document}